\documentclass[12pt]{article}
\usepackage{amsfonts}
\usepackage{amsmath}
\usepackage{amsbsy}
\usepackage{amssymb}
\input{epsf}

\font\grb=eurb10
\def\bphi{\hbox{\grb\char'047}\,}
\def\bpsi{\hbox{\grb\char'040}\,}

\def\vpint{\mathop{\scriptstyle{\mathbf{\diagup}}\hskip-1.9ex \displaystyle{\int}}}

\begin{document}

\title{\bf Monodromy-data parametrization of spaces
of local solutions of integrable reductions of Einstein's
field equations}
\author{
G.~A.~Alekseev
\footnote{E-mail: G.A.Alekseev@mi.ras.ru}
}
\date{}
\maketitle
\begin{abstract}
For the fields depending on two of the four space-time coordinates only, the spaces of local solutions of various integrable reductions of Einstein's field equations are shown to be the subspaces of the spaces of local solutions of the ``null-curvature'' equations constricted by a requirement of a universal (i.e. solution independent) structures of the canonical Jordan forms of the unknown matrix variables. These spaces of  solutions of the ``null-curvature'' equations can be parametrized by a finite sets of free functional parameters -- arbitrary holomorphic (in some local domains) functions of the spectral parameter which can be interpreted as the monodromy data on the spectral plane of the fundamental solutions of associated linear systems. Direct and inverse problems of such mapping (``monodromy transform''), i.e. the problem of finding of the monodromy data for any local solution of the ``null-curvature'' equations with given canonical forms, as well as the existence and uniqueness of such solution for arbitrarily chosen monodromy data are shown to be solvable unambiguously. The linear singular integral equations solving the inverse problems and the explicit forms of the monodromy data corresponding to the spaces of solutions of the symmetry reduced Einstein's field equations are derived.
\end{abstract}


\subsection*{Introduction}
Einstein's field equations for the field configurations that admit a two-dimensional Abelian group of space-time symmetries are well known to be integrable in many physically interesting cases. Many principal questions of the gravitation theory concerning the nonlinearity of gravitational fields and their interactions with matter fields can be analyzed if we restrict our considerations to  stationary axisymmetric fields, or to plane, cylindrical and some other kinds of waves, or to inhomogeneous cosmological models with spacelike symmetries, for which the integrability of the reduced field equations opens new ways for the analytical investigation.

The development of effective methods for solving Einstein's field equations  started with the discovery that the Einstein's equations for gravitational fields in vacuum are integrable if the space-time metric depends only on two coordinates  and has a block-diagonal form \cite{Belinskii-Zakharov:1978} -- \cite{Hauser-Ernst:1979a}. A bit later it was shown that for space-times with this symmetry the integrability features were revealed also for the Einstein - Maxwell equations for gravitational and electromagnetic fields outside their sources \cite{Kinnersley:1977} -- \cite{Alekseev:1980}, for the Einstein - Maxwell - Weyl equations for gravitationally interacting electromagnetic and massless two-component spinor fields \cite{Alekseev:1983}, for the Einstein equations with the energy - momentum tensor of the minimally coupled massless scalar field or for the ideal fluid with the stiff matter equation of state  ($\varepsilon=p$) \cite{Belinskii:1979a}. In addition to these, the cases which were found to be integrable include the similar reductions of the Einstein equations which describe the dynamics of some generalized gravity models, such as, for example, a five-dimensional Kaluza - Klein theory  with a dilaton \cite{Belinskii:1979b} or some string gravity models which arise in the classical (low-energy) limit of the bosonic part of the effective action and describe a specific interaction of various   massless fields -- the dilaton, axion, $U(1)$-gauge, scalar (moduli)  and gravitational fields \cite{Bakas:1994} --  \cite{Alekseev:2004}.

The integrability of the mentioned above reduced field equations allows  developing  of various effective methods for construction of multiparametric families of their solutions with different physical interpretations as well as formulating of some more general approaches to investigating wide classes of solutions.\footnote{The descriptions of many related results can be found in \cite{SKMHH:2003} -- \cite{Sibgatullin:1984}.} In particular, in \cite{Belinskii-Zakharov:1978} it was shown that the constructing vacuum solutions of a general (``nonsoliton'') type can be reduced to solving the matrix Riemann - Hilbert problem whose solution is equivalent to solution of a system of linear singular integral equations on the spectral plane. For a subclass of vacuum and electrovacuum stationary axisymmetric fields with a regular axis of symmetry the effective construction of the internal symmetry transformations for the Einstein - Maxwell equations was reduced to solution of a homogeneouse matrix Hilbert problem and an equivalent system of linear singular integral equations  \cite{Hauser-Ernst:1979b}. Later  this system has been reduced  to a much simpler (non-matrix) linear singular integral equations \cite{Sibgatullin:1984}.\footnote{The simplifying constraint imposed in \cite{Hauser-Ernst:1979a}, \cite{Hauser-Ernst:1979b}, \cite{Sibgatullin:1984} on the space of local solutions under consideration and interpreted for axisymmetric fields as the condition of regularity of the axis of symmetry, halves a number of arbitrary functional parameters in the kernels of the corresponding integral equations. This constraint is justified geometrically for axisymmetric fields, but imposing a similar constraint on fields of a different nature, e.g., colliding plane waves or cosmological models, exclude some physically interesting types of  field configurations from consideration.} For vacuum fields an analogue of the scattering data was suggested and the corresponding discrete form of the  Gelfand - Levitan - Marchenko equation was constructed in \cite{Neugebauer:1981}.

Using the experience of the mentioned above approaches and investigating common properties of many known particular solutions, the author has found a general approach to the analysis of the integrable two-dimensional reductions of Einstein's field equations described above \cite{Alekseev:1985}, \cite{Alekseev:1987}, \cite{Alekseev:2000}. In every case, this approach provides some base for consideration of the entire space of local solutions without any simplifying constraints. Application of this approach allows to calculate new hierarchies of the families of solutions \cite{Alekseev:1993a} and the nonlinear superpositions of some physically interesting types of fields \cite{Alekseev-Garcia:1996}, as well as to consider various initial and boundary value problems \cite{Alekseev:1993b}. This approach called as the ``monodromy transform'' method suggests, following the analogy with the inverse scattering method, to represent the dynamical parts of the reduced field equations as the compatibility conditions of some overdetermined linear system with a free complex (``spectral'') parameter and with additional constraints implying the existence for this system of the matrix integrals of appropriate structures.\footnote{For different mentioned above field equations the dimension of the associated linear system, the character of the singular points and the structures of the matrix integrals can be different.} For each of the integrable cases there exists such associated linear system that all monodromy properties on the spectral plane of its fundamental solution (normalized at some initial space-time point by a unit matrix) are determined by a set of coordinate independent holomorphic (in some local region of the spectral plane) functions of the spectral parameter.

The set of independent functional parameters which arise in this way and which are called as the monodromy data, can play the role similar to the role played by the scattering data in the inverse scattering method. This means that the analytical structures of these functions (the monodromy data) characterizes uniquely each local solution of the field equations, so that there exists a one to one correspondence between the space of all local solutions and a free space of all monodromy data functions. Thus, the direct and inverse problems  arise naturally for such mapping (``monodromy transform''). The problem of finding of the monodromy data for a given local solution of the field equations (the ``direct problem'') reduces to constructing the corresponding normalized fundamental solution of the associated linear system with the uniquely defined (provided the choice of the initial point was made) initial data. A system of linear singular integral equations was found for solution of the ``inverse problem'', i.e. for constructing of the local solution corresponding to arbitrarily chosen monodromy data. The right-hand sides and the scalar kernels of these equations are respectively expressed linearly and bilinearly in terms of the monodromy data and their complex-conjugate functions.
The structure of these integral equations is such that for any choice of the monodromy data functions these equations have a unique (locally defined) solution. For any given monodromy data and for the corresponding solution of these integral equations all of the metric components and field potentials can be expressed in quadratures.

An essential feature of the approach described here is that we solve at first a more general problem of solving some matrix  ``null-curvature'' equations with an additional requirement fixing the universal (i.e., solution-independent) structures of the canonical Jordan forms of the unknown matrix functions. Constructing a general solution of this problem reduces to solving of the mentioned above system of linear singular integral equations, whose structure can be expressed explicitly in terms of a set of free functional parameters (the ``extended monodromy data''). Then the general solution of the reduced Einstein's field equations under consideration can be obtained from this, if we impose on the extended monodromy data some simple algebraic constraints, solvable explicitly. Because of these constraints, this solution is determined by half the number of arbitrary functional parameters which we called simply as the ``monodromy data''. In the previous author's papers the construction described above was presented very schematically for the most simple cases of vacuum Einstein's equations and electrovacuum Einstein - Maxwell field equations only. Therefore, the main purpose of the present paper is to describe this construction in more detail also including some important features of the other integrable reductions of Einstein's field equations.

\subsection*{Integrable reductions of Einstein's field equations}

In the four-dimensional space-time admitting two-dimensional Abelian isometry group the local coordinates can be chosen in such a way that all of the metric components and field potentials depend on only two of the four coordinates $x^\mu=\{x^1,x^2\}$ but not on the coordinates $x^a=\{x^3,x^4\}$. The reductions of the field equations corresponding to this symmetry describe physically different types of fields which depend either only on time and one spatial coordinate (the ``hyperbolic'' case) or only on two spatial coordinates (the ``elliptic''  case). It is convenient to consider both cases in a unified form, where these cases will be distinguished by the sign symbol $\epsilon=\pm1$ and its ``square root'' $j$, so that $\epsilon=1$ and $j=1$ for the hyperbolic reductions, and $\epsilon=-1$ and $j=i$ for the elliptic ones.

\paragraph{The choice of coordinates.}
For the local coordinates $x^\mu$ it is convenient to use the so called generalized Weyl coordinates $\alpha$ and $\beta$, which are determined by the space-time geometry itself. The function $\alpha(x^1,x^2)$ determines the area measure on the two-dimensional orbits of the isometry group, and in accordance with the field equations the function $\alpha(x^1,x^2)$ is ``harmonic'', while the function  $\beta(x^1,x^2)$ is defined as its ``harmonic'' conjugated one:
$$d\,\,{}^\ast\! d\alpha=0,\qquad d\beta=-\epsilon\, {}^\ast\! d \alpha$$
where $d$ denotes the external derivative and ${}^\ast$ denotes its dual conjugation. Hence, $d\,{}^\ast\! d$ in the conformal coordinates reduces to the two-dimensional Laplace operator in the elliptic case or to the d'Alembert operator  in the hyperbolic case. In some cases (for example, in the problem of colliding plane waves or when considering other initial or boundary value problems) it is convenient to use other coordinates, and  $\alpha(x^1,x^2)$ then becomes an additional dynamical variable. In what follows, we use for coordinates the linear combinations
$$\xi=\beta+j\alpha,\qquad \eta=\beta-j\alpha,$$
which are two real null coordinates (``light cone'' variables) in the hyperbolic case or two mutually complex conjugate coordinates in the elliptic case.

\paragraph{Reduced dynamical equations.}

The reduced dynamical parts of the  Einstein equations for the vacuum or of the electrovacuum Einstein-Maxwell field equations, and of the Einstein - Maxwell - Weyl field equations can be presented in the rspective forms of the Ernst equations \cite{Ernst:1968} or generalized Ernst equations \cite{Alekseev:2000} for the complex scalar functions -- the Ernst potentials  ${\cal E}(x^1,x^2)$ and $\Phi(x^1,x^2)$. These equations are most compact written in terms of differential forms (for simplicity, we omit the symbol $\wedge$ for the external multiplication of these forms):
\begin{equation}\label{ErnstEqsA}
\left\{\begin{array}{l} d\,{}^{\ast}d{\cal
E}+\displaystyle{d(\alpha+i\delta)\over\alpha}\,{}^{\ast}d{\cal
E} -\displaystyle{(d {\cal E}+2\overline{\Phi} d\Phi) \over
\mbox{Re\,}{\cal E}+\Phi \overline{\Phi}}\,{}^{\ast}d{\cal E}
=0\\[1em]
d\,{}^{\ast}d\Phi+\displaystyle{d(\alpha+i\delta)\over\alpha}\,
{}^{\ast}d\Phi -\displaystyle{(d {\cal E}+2\overline{\Phi} d\Phi)
\over \mbox{Re\,}{\cal E}+\Phi
\overline{\Phi}}\,{}^{\ast}d\Phi=0\\[1.2em]
d\,{}^{\ast}d\alpha=0,\quad\qquad d\beta \equiv-\epsilon\,
{}^{\ast}d\alpha\\[1ex]
d\,{}^{\ast}d\gamma=0,\quad\qquad d\delta \equiv {}^{\ast}d\gamma
\end{array}\right.
\end{equation}
The spinor field here is represented by a real ``harmonic'' function  $\gamma(x^1,x^2)$. This function is interpreted as a potential for the components of the neutrino vector current. The linear equation for this function can be solved explicitly and its general solution is a sum of two arbitrary real functions (in the hyperbolic case) or two mutually complex-conjugate functions (in the elliptic case), of which one depends only  on $\xi$ and the other, only on $\eta$. For electrovacuum fields,  $\gamma\equiv 0$. For vacuum fields,  we also have $\Phi\equiv 0$; therefore, only the Ernst equation for one complex potential ${\cal E}(x^1,x^2)$ and the linear equation for  $\alpha$ pertain in this case. The presence of a minimally coupled scalar field, as well as of a perfect fluid with the stiff matter equation of state, do not affect the reduced dynamical equations and are manifested only in the constraint equations. The solution of these constraint equations can be calculated in quadratures for every solution of the dynamical equations. The reduced equations, which describe the dynamics of bosonic fields in the low-energy limits of some string gravity models and which include the axion, dilaton, Abelian gauge vector fields and the scalar (moduli) fields, can also take the forms of integrable matrix analogues of the vacuum Ernst equation \cite{Bakas:1994} -- \cite{Gal'tsov-et-alii},\cite{Alekseev:2004}:
\begin{equation}\label{ErnstEqsB}
\left\{\begin{array}{l} d\,{}^{\ast}d{\cal
E}+\displaystyle{d\alpha\over\alpha}\,{}^{\ast}d{\cal
E} -d {\cal E}\cdot(\hbox{Re}\,{\cal E})^{-1}\cdot
\,{}^{\ast}d{\cal E}=0\\[1em]
d\,{}^{\ast}d\alpha=0,
\end{array}\right.
\end{equation}
where ${\cal E}(x^1,x^2)$ is a Hermitian  or complex symmetric  $d\times d$ - matrix function.

\paragraph{``Null-curvature'' representation of the integrable reductions of Einstein's field equations.}
The dynamical equations (\ref{ErnstEqsA}) or (\ref{ErnstEqsB}) can be presented in the form of a system of the first-order equations (with a subsequent reduction) for complex $N\times N$ matrices ${\bf U}$ and ${\bf V}$, where $N=2$ for vacuum fields, $N=3$ for the Einstein - Maxwell and Einstein - Maxwell - Weyl fields and $N=2d$ ($d>1$) for the string gravity models which lead to the equations (\ref{ErnstEqsB}). These matrix equations include the equations which are similar to the known ``null-curvature'' equations and additional constraints requiring the canonical (Jordan) forms of the unknown matrix coefficients ${\bf U}$ and ${\bf V}$ to have universal (i.e., solution independent) diagonal forms with constant components if the Weyl spinor field is absent or with components depending on the potential of the Weyl spinor field if it is present \cite{Alekseev:1980}, \cite{Alekseev:1983}, \cite{Alekseev:2000}, \cite{Alekseev:2004}:
\begin{equation}\label{NullCurv}
\left\{\begin{array}{l}{\bf U}_\eta+{\bf
V}_\xi+\displaystyle\frac 1{i (\xi-\eta)} [{\bf U},{\bf V}] =
0\\[2ex]
{\bf U}_\eta-{\bf V}_\xi=0
\end{array}\qquad\right\Vert\qquad
\begin{array}{l}
{\bf U}={\cal F}_+ {\bf U}_{(o)}{\cal F}_+^{-1}\\[2ex]
{\bf V}={\cal F}_- {\bf V}_{(o)}{\cal F}_-^{-1}
\end{array}
\end{equation}
where ${\cal F}_\pm$ are the matrices transforming  ${\bf U}(\xi,\eta)$ and ${\bf V}(\xi,\eta)$ to their canonical forms ${\bf U}_{(o)}$ and ${\bf V}_{(o)}$,
\begin{equation}\label{CanonicalForms}
\begin{array}{ll}
{\bf U}^{(N=2)}_{(o)}=\hbox{diag}\,\{i,\,0\},&
{\bf V}^{(N=2)}_{(o)}=\hbox{diag}\,\{i,\,0\},\\[1ex]
{\bf U}^{(N=3)}_{(o)}=\hbox{diag}\,\{i+\mu_+(\xi),\,0,\,0\},&
{\bf V}^{(N=3)}_{(o)}=\hbox{diag}\,\{i+\mu_-(\eta),\,0,\,0\},\\[1ex]
{\bf U}^{(N=2d)}_{(o)}=\hbox{diag}\, \{\underset d{\underbrace{i,\ldots,i}},\underset d{\underbrace{0,\ldots,\,0}}\},&
{\bf V}^{(N=2d)}_{(o)}=\hbox{diag}\, \{\underset d{\underbrace{i,\ldots,i}},\underset d{\underbrace{0,\ldots,\,0}}\}.
\end{array}
\end{equation}
As it was mentioned above, $N=2$ for vacuum fields; the presence of electromagnetic fields increases the dimensions of these matrices to $N=3$, and the presence of the spinor field leads to a dependence of the eigenvalues of these matrices on the components of the current vector of the neutrino field: $\mu_+(\xi)=2\partial_\xi\gamma$ and $\mu_-(\eta)=2\partial_\eta\gamma$. The functions $\mu_+(\xi)$ and $\mu_-(\eta)$ must be real in the hyperbolic case; in the elliptic case these functions, being the derivatives of a real harmonic function, must satisfy the relation  $\mu_+(\overline{\zeta})=\overline{\mu_-(\zeta)}$. The case $N=2d$ with $d>1$ corresponds to the string theory models described by the matrix equations (\ref{ErnstEqsB}). The case $d=1$ coincides with the case $N=2$, but the last case is mentioned here and below separately because of its classical interpretation.

To provide the relations between the components of ${\bf U}$, ${\bf V}$ (or the components of ${\cal F}_\pm$  in (\ref{NullCurv})) and the field variables and their first derivatives which reduce   (\ref{NullCurv}), (\ref{CanonicalForms}) to the Ernst equations (\ref{ErnstEqsA}) or (\ref{ErnstEqsB}), the components of ${\bf U}$ and ${\bf V}$ must satisfy additional constraints. These constraints are presented below in terms of restrictions on the fundamental solutions of the associated linear systems with the spectral parameter, whose compatibility conditions coincide with (\ref{NullCurv}).

\subsection*{Spectral problems for the equations (\ref{NullCurv}),
(\ref{CanonicalForms}) and (\ref{ErnstEqsA}), (\ref{ErnstEqsB})}
The equations (\ref{NullCurv}) are the compatibility conditions for the  overdetermined $N\times N$ linear system ($N=2,3$ or $2 d$)
\begin{equation}\label{LinSys}
\left\{\begin{array}{l}
2 i(w-\xi)\partial_\xi{\bf \Psi}={\bf U}(\xi,\eta)\,{\bf \Psi}\\[1ex]
2 i(w-\eta)\partial_\eta{\bf \Psi}={\bf V}(\xi,\eta)\,{\bf \Psi}
\end{array}\right.
\end{equation}
where $w$ is a complex (spectral) parameter. For vacuum and electrovacuum fields this system was derived by different authors and it was used for different constructions in \cite{Kinnersley-Chitre:1978} -- \cite{Alekseev:1980}. The additional constraints described below, which are necessary and sufficient for equivalence of the entire spectral problem to the Ernst equations, were found in  \cite{Alekseev:1985}.

\paragraph{The spectral problem for  ``null-curvature'' equations  (\ref{NullCurv}) with constraints (\ref{CanonicalForms}).}
The equations (\ref{NullCurv}), (\ref{CanonicalForms}) are equivalent to the spectral problem for the three complex  $N\times N$-matrix functions
\begin{equation}\label{MatricesA}
{\bf \Psi}(\xi,\eta,w),\quad {\bf U}(\xi,\eta),\quad{\bf V}(\xi,\eta)
\end{equation}
which satisfy (\ref{LinSys}) and whose canonical Jordan forms  ${\bf U}_{(o)}$ and ${\bf V}_{(o)}$ have the form  (\ref{CanonicalForms}). Just the solution of this spectral problem will be interesting for us first of all, because this problem can be reduced to the spectral problem for the generalized Ernst equations, if we impose on it's solutions some additional constraints providing the existence for each of the systems (\ref{LinSys}) of matrix integrals of some special structures.

\paragraph{The spectral problem for generalized Ernst equations (\ref{ErnstEqsA}) and (\ref{ErnstEqsB}).}
For the Ernst equations (\ref{ErnstEqsA}) and (\ref{ErnstEqsB}) the spectral problems can be formulated very similarly to each other as the problems of finding of the four complex $N\times N$-matrix functions
\begin{equation}\label{MatricesB}
{\bf \Psi}(\xi,\eta,w),\quad {\bf U}(\xi,\eta),\quad{\bf V}(\xi,\eta),\quad{\bf W}(\xi,\eta,w)
\end{equation}
where the first three matrices must satisfy to the system (\ref{LinSys}), and the matrices ${\bf U}$ and ${\bf V}$ must have the canonical Jordan forms (\ref{CanonicalForms}). An additional condition of this spectral problem is that each of the systems  (\ref{LinSys}) must have the matrix integrals of special structures. In the cases of scalar Ernst potentials ($N=2,3$) or in the matrix case ($N=2 d$, $d>1$) with complex symmetric Ernst potentials these are the Hermitian matrix integrals ${\bf K}(w)$ whose structures include the matrix function  ${\bf W}$ which must be a linear function of the spectral parameter with the constant coefficient ${\bf \Omega}$
\begin{equation}\label{KIntegral}
{\bf \Psi}^\dagger\cdot {\bf W}\cdot{\bf \Psi}={\bf K}(w), \qquad {\bf
K}^\dagger(w)={\bf K}(w),\qquad\displaystyle\frac {\partial {\bf
W}}{\partial w}=4 i{\bf \Omega}.
\end{equation}
Here ${}^\dagger$ denotes Hermitian conjugation, such that $\mathbf{\Psi}^\dagger(\xi,\eta,w)=
\overline{\Psi^T(\xi,\eta,\overline{w})}$.
The matrices ${\bf \Omega}$ are constant and have the structures:
\begin{equation}\label{Omegas}
{\bf \Omega}^{(N=2)}=\left(\hskip-1.5ex\begin{array}{rr}
0&1\\-1&0
\end{array}\right),
\hskip2ex
{\bf \Omega}^{(N=3)}=\left(\hskip-1.5ex\begin{array}{rrr}
0&1&0\\-1&0&0\\0&0&0 \end{array}\right),
\hskip2ex
{\bf \Omega}^{(N=2d)}=\left(\hskip-1.5ex\begin{array}{rr}
0&I_d\\-I_d&0
\end{array}\right)
\end{equation}
where $I_d$ is $d\times d$ unit matrix.
In the matrix case ($N=2d$, $d>1$) we require also the existence of another matrix integral ${\bf L}(w)$ which is antisymmetric in the case of complex symmetric matrix Ernst potentials (\,\,${}^T$ denotes the transposition of matrices):
\begin{equation}\label{LIntegral}
S\,\,\,\,{\bf \Psi}^T\cdot {\bf \Omega}\cdot{\bf \Psi}={\bf L}(w)\qquad {\bf L}^T(w)=-{\bf L}(w)\qquad
S^2\equiv (w-\xi)(w-\eta)
\end{equation}
For another matrix case ($N=2d$, $d>1$) with Hermitian, not symmetric, matrix Ernst potentials, the matrix integral ${\bf K}(w)$ must be complex symmetric, and the matrix integral ${\bf L}(w)$ must be Hermitian. In this case, the matrix $\mathbf{\Omega}$  is symmetric: $$\mathbf{\Omega}=\left(\hskip-1.5ex\begin{array}{rr}
0&I_d\\I_d&0
\end{array}\right)$$
(see \cite{Alekseev:2004} for more details). The subsequent considerations of the cases of Hermitian and symmetric matrix Ernst potentials can proceed very similarly, and we consider only the symmetric case in detail for simplicity.

\paragraph{Equivalence of the spectral problem  (\ref{CanonicalForms}), (\ref{LinSys}), (\ref{KIntegral})-- (\ref{LIntegral}) to the Ernst equations.}
It is easy to see from (\ref{KIntegral}) that  ${\bf W}$ can be written as
$$
{\bf W}=4 i(w-\beta) {\bf \Omega}+{\bf G}(\xi,\eta).
$$
where $\beta=(\xi+\eta)/2$, and the matrix ${\bf G}$ is Hermitian. Then for each $N=2,3$ or in the symmetric matrix case with $N=2 d$ the conditions (\ref{CanonicalForms}), (\ref{LinSys}), (\ref{KIntegral})-- (\ref{LIntegral}) allow  expressing the components of ${\bf G}$ in terms of a smaller set of functions, and expressing the components of ${\bf U}$ and ${\bf V}$ in terms of these functions and their first derivatives. These functions can be identified with the field variables and the conditions  (\ref{CanonicalForms}), (\ref{LinSys}), (\ref{KIntegral})-- (\ref{LIntegral}) for them reduce to the  Ernst equations (\ref{ErnstEqsA}) or (\ref{ErnstEqsB}).

\paragraph{Normalization of solutions.} The spectral problem
(\ref{CanonicalForms}), (\ref{LinSys}) as well as  (\ref{CanonicalForms}), (\ref{LinSys}), (\ref{KIntegral})-- (\ref{LIntegral}) admit obvious gauge transformations. These  consist of two commuting  groups of transformations. One of them is a group of linear transformations of the form
\begin{equation}\label{Atransform}
\left.\begin{array}{lcl}
\mathbf{U}\to \mathbf{A}\cdot\mathbf{U}\cdot\mathbf{ A}^{-1}&&
\mathbf{\Psi}\to
\mathbf{A}\cdot\mathbf{\Psi}\\[1ex]
\mathbf{V}\to \mathbf{A}\cdot\mathbf{V}\cdot \mathbf{A}^{-1}&&\mathbf{W}\to
(\mathbf{A}^{-1})^\dagger\cdot\mathbf{W}\cdot \mathbf{A}^{-1}
\end{array}\quad\right\Vert\quad
\begin{array}{l}
\mathbf{A}^\dagger\cdot \mathbf{\Omega}\cdot \mathbf{A}=\mathbf{\Omega}\\[1ex]
\mathbf{A}^T\cdot \mathbf{\Omega}\cdot \mathbf{A}=\mathbf{\Omega}
\end{array}
\end{equation}
where the constant matrix $\mathbf{A}$ must satisfy the relations on the right\footnote{In each particular case it is easy to find a general forms of these matrices. In particular, for $N=2$ the matrix $\mathbf{A}$ must be real and unimodular, while for the case $N=3$, in which $\mathbf{\Omega}$  degenerates, $\mathbf{A}$ can be complex and has more complicated structure.}. Another group consists of linear transformations of the fundamental solutions of the linear systems which affect the values of the matrix integrals:
\begin{equation}\label{Ctransform}
\mathbf{\Psi} (\xi,\eta,w)\to \mathbf{\Psi} (\xi,\eta,w)\cdot\mathbf{C} (w)\quad\left\Vert\quad
\begin{array}{l}
{\bf K}(w)\longrightarrow {\bf C}^\dagger(w)\cdot{\bf K}(w)\cdot{\bf C}(w)\\ {\bf L}(w)\,\longrightarrow {\bf C}^T(w)\cdot{\bf L}(w)\cdot{\bf C}(w)
\end{array}\right.
\end{equation}
The transformations (\ref{Atransform}) correspond to the Killing vector linear transformations, to constant shifts of the values of the electromagnetic potentials, to the dual transformations of the electromagnetic fields, and so on. These transformations can be used to reduce the values of all field variables at a chosen initial point $P_0(\xi_0,\eta_0)$ to some ``standard'' initial values, which can coincide, for example, with their values for the Minkowski space-time. This determines also the initial value of the matrix function $\mathbf{W}$:   \begin{equation}\label{Wo}
\mathbf{W}_0(w)\equiv \mathbf{W}(\xi_0,\eta_0,w)
=4 i(w-\beta_0)\mathbf{\Omega}+\mathbf{G}(\xi_0,\eta_0).
\end{equation}
We use the transformations (\ref{Ctransform}) with appropriate  $\mathbf{C}(w)$ to normalize the value of $\mathbf{\Psi}$ at the initial point by a unit matrix
\begin{equation}\label{NormPsi}
\mathbf{\Psi} (\xi,\eta,w)\to \mathbf{\Psi} (\xi,\eta,w)\cdot\mathbf{\Psi}^{-1}(\xi_0,\eta_0,w)\qquad
\Longrightarrow
\qquad
\mathbf{\Psi}(\xi_0,\eta_0,w)\equiv \mathbf{I}
\end{equation}
because just this choice leads to the most simple analytical properties of $\mathbf{\Psi}$ on the spectral plane. This normalization determines also the values of the matrix integrals $\mathbf{K}(w)$ and $\mathbf{L}(w)$, and the conditions (\ref{KIntegral}), (\ref{LIntegral}) take the forms
\begin{equation}\label{NormIntegrals}
{\bf \Psi}^\dagger\cdot {\bf W}\cdot{\bf \Psi}={\bf W}_0(w),\qquad
\lambda_+\lambda_-\, {\bf \Psi}^T \cdot{\bf \Omega}\cdot{\bf \Psi}={\bf \Omega}, \qquad\displaystyle\frac {\partial {\bf
W}}{\partial w}=4 i{\bf \Omega}
\end{equation}
where $\lambda_+^2=(w-\xi)/(w-\xi_0)$, $\lambda_-^2=(w-\eta)/(w-\eta_0)$ and  $\lambda_\pm(w=\infty)=1$.

We consider now the properties of a general solution of the first of the spectral problems described above without any further simplifying assumptions.

\subsection*{The spaces of local solutions}
Let us consider a two-dimensional real manifold  $\mathcal{M}^2$ with coordinates $(\xi,\eta)$ and chose some ``initial'' point  $P_0(\xi_0,\eta_0)$ in it. Because  we consider only local solutions, we take the manifold to be a local neighborhood of the point  $P_0$ in $\mathbb{R}^2$. Although the coordinates $\xi$ and $\eta$ by  definition are real in the hyperbolic case or mutually complex conjugate in the elliptic case, it is convenient to extend their domains and to take $\xi$ and $\eta$ to be two independent complex variables running some local region $\Omega^2$ near the point $P_0$ in $\mathbb{C}^2$ in both cases. We introduce the following definitions.

\textsc{Definition 1.} \textit{A pair of matrix functions  $\mathbf{U}(\xi,\eta)$, $\mathbf{V}(\xi,\eta)$, that are holomorphic in some neighborhood of a given initial point  $P_0(\xi_0,\eta_0)$, that satisfy the equations (\ref{NullCurv}) there, and that have canonical Jordan forms (\ref{CanonicalForms}) at every point in this neighborhood, is called here a local solution at the point $P_0$ of the ``null-curvature'' equations with the Jordan condition (\ref{CanonicalForms}). All such solutions constitute the space of local solutions of these equations at this point.}

\textsc{Definition 2.} \textit{A complete set of the field variables  (the metric components, the potentials of matter fields, and the corresponding Ernst potentials) that take at the initial point $P_0(\xi_0,\eta_0)$ some ``standard'' values, that are holomorphic functions of $\xi$, $\eta$ in some neighborhood of this point, and that satisfy a complete set of the field equations (in the form of the Ernst equations (\ref{ErnstEqsA}) or (\ref{ErnstEqsB}), for example) in this  neighborhood is called a (normalized) local solution of the field equations at the point $P_0$. All such solutions constitute the space of (normalized) local solutions of the field equations at this point.}\\[0ex]

Without any loss of generality we assume that the neighborhood $\Omega^2$ of the initial point $P_0$ mentioned in the above definitions takes the form $\Omega_{\xi_0}\times\Omega_{\eta_0}$, where $\Omega_{\xi_0}$ and $\Omega_{\eta_0}$ are the local domains of complex variables $\xi$ and $\eta$  which they range near their initial values $\xi_0$ and $\eta_0$ respectively. The regions $\Omega_{\xi_0}$ and $\Omega_{\eta_0}$ are chosen such that their images on the spectral plane \begin{equation}\label{Omegapm}
\Omega_+=\{w\,\vert\, w=\xi\in\Omega_{\xi_0}\},\qquad
\Omega_-=\{w\,\vert\, w=\eta\in\Omega_{\eta_0}\}
\end{equation}
are convex, non-intersecting, and symmetric with respect to the real axis in the hyperbolic case or symmetric to each other with respect to this axis in the elliptic case.\footnote{We recall that in the domain of regularity of solutions which we consider here the function $\alpha$ is positive by its definition, and therefore, in the elliptic case the points of $\Omega_+$ are located above the real axis only, and the points of $\Omega_-$ are located below this axis. The points  $\alpha=0$, which may correspond to various possible singularities of the solution geometry (such as curvature singularities or the Killing and Cauchy horizons), are excluded from our consideration by an appropriate choice of $\Omega^2$ as a small enough neighborhood of the initial point where $\alpha_0\ne 0$ in accordance with the normalization conditions. However, in some solutions, the points  $\alpha=0$ correspond to ``removable'', pure coordinate singularities, such as the points of the axis of symmetry for stationary axisymmetric fields or cylindrical waves. For these cases it may be convenient to chose the initial point on the axis of symmetry. Then the regions $\Omega_+$ and $\Omega_-$ intersect. In this case, the known condition of regularity of the axis of symmetry reduces the space of local solutions in such a way that an ambiguity of solutions do not arise and all our constructions (with the additional constraints taken into account) can be used for this subclass of solutions as well.}
Thus, any local solution ${\bf U}$, ${\bf V}$ of the equations  (\ref{NullCurv}) with the constraints (\ref{CanonicalForms}) is holomorphic in some neighborhood $\Omega^2$ of the initial point:
$${\bf U}, {\bf V}:\quad \Omega^2\longrightarrow GL(N,\mathbb{C}),\qquad
\Omega^2=\Omega_{\xi_0}\times\Omega_{\eta_0}\hskip2ex \subset \,\mathbb{C}^2
$$
where $N=2,3$ or $N=2 d$ with $d>1$, in accordance with the dimensions of the linear systems. Therefore, we have to consider the analytical properties of the corresponding fundamental solution ${\bf \Psi}(\xi,\eta,w)$ of the linear system (\ref{LinSys}), normalized by the condition  (\ref{NormPsi}), in the region $\Omega^3$:
$${\bf \Psi}:\quad \Omega^3\longrightarrow GL(N,\mathbb{C}),\qquad
\Omega^3=\Omega^2\times\overline{{\mathbb{C}}}\hskip4ex \subset \,\overline{\mathbb{C}}^3
$$
where $\xi$ and $\eta$ run the domains $\Omega_{\xi_0}$ and $\Omega_{\eta_0}$ near the initial points $\xi_0$ and $\eta_0$ respectively and $w$ takes its values in the extended complex plane $\overline{\mathbb{C}}$.

\subsection*{Analytical structur of $\mathbf{\Psi}(\xi,\eta,w)$}
\paragraph{The singular points of $\mathbf{\Psi}(\xi,\eta,w)$ and $\mathbf{\Psi}^{-1}(\xi,\eta,w)$.}
In the region $\Omega^3$, the singular points of the linear system (\ref{LinSys}) are represented by the parts of complex hyperplanes
$\Omega^3 \cap \{w=\xi\}$ and $\Omega^3 \cap \{w=\eta\}$. The normalization (\ref{NormPsi}) of the solution  at the initial point introduces two additional sets of the singular points represented by the parts of complex hyperplanes  $\Omega^3 \cap \{w=\xi_0\}$ and $\Omega^3 \cap \{w=\eta_0\}$. In accordance with the known theorems of the analytical theory of ordinary differential equations, the functions $\mathbf{\Psi}(\xi,\eta,w)$ and $\mathbf{\Psi}^{-1}(\xi,\eta,w)$,  are (multivalued) analytical functions in the none-simply connected region
represented by the part of $\Omega^3$ outside its intersection with the mentioned just above four complex hyperplanes.\footnote{General analytical properties of these functions, including their dependence on the choice of the point of normalization, have been described in detail in the paper of Hauser and Ernst \cite{Hauser-Ernst:1980}. In this paper the authors had described also the properties of the holomorphic branch of $\mathbf{\Psi}(\xi,\eta,w)$ near the regular points of the axis of symmetry using a cut of the special structure appropriate only for a class of stationary axisymmetric fields with a regular axis of symmetry. Below we consider the cut of another structure which can be used for any local solution without any additional constraints on the class of fields. The properties of the holomorphic branch of $\mathbf{\Psi}(\xi,\eta,w)$ corresponding to this cut had been described and used in the author's papers \cite{Alekseev:1985}, \cite{Alekseev:1987}. Later in the paper of Hauser and Ernst \cite{Hauser-Ernst:2001} these properties had been considered in detail for solutions with finite smoothness for the hyperbolic case (the colliding plane waves) and used there for the proofs of the general theorems of existence and uniqueness of solutions of the characteristic initial value problem and for the proof of the Geroch conjecture for hyperbolic Ernst equations. In what follows we adapt the approach
\cite{Alekseev:1985}, \cite{Alekseev:1987}.}

\paragraph{The region of holomorphicity of $\mathbf{\Psi}(\xi,\eta,w)$ and $\mathbf{\Psi}^{-1}(\xi,\eta,w)$.}
To construct the single-valued branch of $\mathbf{\Psi}(\xi,\eta,w)$ in the region $\Omega^3$, it is necessary to make there a cut which would represent  a hypersurface in the real six-dimensional space corresponding to the three dimensional complex region $\Omega^3$. This cut should make this region simply connected and include all four mentioned earlier sets $\Omega^3 \cap \{w=\xi\}$, $\Omega^3 \cap \{w=\eta\}$ and $\Omega^3 \cap \{w=\xi_0\}$, $\Omega^3 \cap \{w=\eta_0\}$.
We construct this cut as a connected hypersurface  $\widetilde{\cal L}$, consisting of three ruled surfaces: $\widetilde{\cal L}={\cal L}_+\cup {\cal L}_0 \cup {\cal L}_-$, where each of the hypersurfaces  ${\cal L}_+$, ${\cal L}_0$ and ${\cal L}_-$ consists of the straight lines which join on each plane $\{\xi=\hbox{const},\eta=\hbox{const}\}$ the points $w=\xi_0$ and $w=\xi$,  the points $w=\xi_0$ and $w=\eta_0$, and, at last, the points $w=\eta_0$ and $w=\eta$ respectively. If necessary, this hypersurface can be transformed to a smooth one using an appropriate continuous deformation. The region $\Omega^3\backslash \widetilde{\mathcal{L}}$ obtained in this way is simply connected. This is so because any section  of the cut $\widetilde{\mathcal{L}}$ by the plane $\{\xi=\hbox{const}, \eta=\hbox{const}\}$ is represented on the spectral plane $w$ by some continuous curve $\widetilde{L}$, which consists of three curves: $\widetilde{L}=L_+\cup L_0 \cup L_-$. The part of the Riemann  sphere outside this curve is simply connected, and therefore, the region
$\Omega^3\backslash \widetilde{\mathcal{L}}$ is also simply connected.
Therefore, in the region $\Omega^3\backslash \widetilde{\mathcal{L}}$ the normalized function $\mathbf{\Psi}(\xi,\eta,w)$ possesses a holomorphic branch.

We show now that on the intermediate part ${\cal L}_0$ of the cut described above this branch is continuous. Therefore, the cut  ${\cal L}_0$ is not necessary and its points are also the points of holomorphicity of $\mathbf{\Psi}(\xi,\eta,w)$. To show this, we use for the holomorphic branch of the normalized fundamental solution $\mathbf{\Psi}(\xi,\eta,w)$ constructed here the factorizations of the forms\footnote{These factorizations exist because we can take the limits  $\eta\to\eta_0$ and $\xi\to\xi_0$ in the first and the second equations  (\ref{LinSys}) respectively using the holomorphicity of  $\mathbf{\Psi}(\xi,\eta,w)$ in  $\Omega^3\backslash \widetilde{\mathcal{L}}$ with respect to all its arguments.}:
\begin{equation}\label{Factorization}
\left.\begin{array}{l}
\mathbf{\Psi}(\xi,\eta,w)=\mathbf{\chi}_+(\xi,\eta,w)\cdot \mathbf{\Psi}_+(\xi,w)\\[2ex]
\mathbf{\Psi}(\xi,\eta,w)=\mathbf{\chi}_-(\xi,\eta,w)\cdot \mathbf{\Psi}_-(\eta,w)
\end{array}\quad\right\Vert\quad
\begin{array}{l}
\mathbf{\Psi}_+(\xi,w)=\mathbf{\Psi}(\xi,\eta_0,w)\\[1ex]
\mathbf{\Psi}_-(\eta,w)=\mathbf{\Psi}(\xi_0,\eta,w)
\end{array}
\end{equation}
where $\mathbf{\Psi}_+$ and $\mathbf{\chi}_+$ are the fundamental solutions of the linear systems
\begin{equation}\label{LinSysPlus}
\left\{\begin{array}{l}
2 i(w-\xi)\partial_\xi \mathbf{\Psi}_+=\mathbf{U}(\xi,\eta_0) \cdot \mathbf{\Psi}_+\\[1ex]
\mathbf{\Psi}_+(\xi_0,w)=\mathbf{I}
\end{array}\right.\qquad
\left\{\begin{array}{l}
2 i(w-\eta)\partial_\eta \mathbf{\chi}_+=\mathbf{V}(\xi,\eta) \cdot \mathbf{\chi}_+\\[1ex]
\mathbf{\chi}_+(\xi,\eta_0,w)=\mathbf{I}
\end{array}\right.
\end{equation}
and $\mathbf{\Psi}_-$ and $\mathbf{\chi}_-$ are the fundamental solutions of the linear systems
\begin{equation}\label{LinSysMinus}
\left\{\begin{array}{l}
2 i(w-\eta)\partial_\eta \mathbf{\Psi}_-=\mathbf{V}(\xi_0,\eta) \cdot \mathbf{\Psi}_-\\[1ex]
\mathbf{\Psi}_-(\eta_0,w)=\mathbf{I}
\end{array}\right.\qquad
\left\{\begin{array}{l}
2 i(w-\xi)\partial_\xi \mathbf{\chi}_-=\mathbf{U}(\xi,\eta) \cdot \mathbf{\chi}_-\\[1ex]
\mathbf{\chi}_-(\xi_0,\eta,w)=\mathbf{I}
\end{array}\right.
\end{equation}
Restricting our consideration for definiteness by the first of these factorizations only, we note that the structure of the systems (\ref{LinSysPlus}) allow us to conclude that in the region  $\Omega^3$ the matrix function $\mathbf{\Psi}_+(\xi,w)$ has the singular points $\Omega^3\cap \{w=\xi\}$ and $\Omega^3\cap \{w=\xi_0\}$, and the function $\mathbf{\chi}_+(\xi,\eta,w)$ has the singular points $\Omega^3\cap \{w=\eta\}$ and $\Omega^3\cap \{w=\eta_0\}$. Therefore, the function $\mathbf{\Psi}_+(\xi,w)$ has a holomorphic branch in the region $\Omega^3\backslash \mathcal{L}_+$, and the function   $\mathbf{\chi}_+(\xi,\eta,w)$ has a holomorphic branch in the region $\Omega^3\backslash \mathcal{L}_-$. This means that the product (\ref{Factorization}) of these functions is also holomorphic on $\mathcal{L}_0$. Thus, we conclude that the normalized fundamental solution of the associated linear system has a holomorphic branch in the region $\Omega^3\backslash \mathcal{L}$ where the cut $\mathcal{L}$
consists of two disconnected parts: $\mathcal{L}=\mathcal{L}_+\cup \mathcal{L}_-$. For simplicity, everywhere below $\mathbf{\Psi}(\xi,\eta,w)$ will denote just this holomorphic branch determined by the normalization condition $\mathbf{\Psi}(\xi_0,\eta_0,w)=\mathbf{I}$.

An important property of the holomorphic branch $\mathbf{\Psi}(\xi,\eta,w)$ constructed above is that $\mathbf{\Psi}(\xi,\eta,w=\infty)=\mathbf{I}$. One can observe this from the structure of the linear system (\ref{LinSys}), which implies in the limit $w=\infty$ that $\partial_\xi\mathbf{\Psi}=0$ and $\partial_\eta\mathbf{\Psi}=0$. In accordance with the normalization conditions, for $w=\infty$ the coordinate independent value of  $\mathbf{\Psi}$ should coincide with the unit matrix.

The matrix function $\mathbf{\Psi}(\xi,\eta,w)$ in the region
$\Omega^3\backslash \mathcal{L}$ has a holomorphic inverse matrix. To see this, we should multiply each of the equations (\ref{LinSys}) by $\mathbf{\Psi}^{-1}$ and take the trace of both their sides. In view of the identity $\hbox{Tr}\,(d\mathbf{\Psi}\cdot \mathbf{\Psi}^{-1})\equiv d(\log\det \mathbf{\Psi})$, both of these  relations can be easily integrated. Taking into account the normalization conditions (\ref{NormPsi}), we can determine the integration constant which is a function of $w$. This leads to the following expression for the determinant of $\mathbf{\Psi}(\xi,\eta,w)$
$$
\det \Vert\mathbf{\Psi}(\xi,\eta,w)\Vert =
\left\{\begin{array}{ll}
\lambda_+^{-1}\lambda_-^{-1}
e^{-i(\sigma_++\sigma_-)}&\hbox{for}\quad N=2,3,\\
\lambda_+^{-d}\lambda_-^{-d}&\hbox{for}\quad N=2 d,\end{array}\right.
$$
where $\lambda_\pm(\xi,\eta,w)$ and $\sigma_\pm(\xi,\eta,w)$ possess the expressions
\begin{equation}\label{Lambdasigma}\begin{array}{lcl}
\lambda_+(\xi,w)=\sqrt{\dfrac{w-\xi}{w-\xi_0}},\quad
\lambda_+(\xi,w=\infty) =1,&&
\sigma_+(\xi,w)=\displaystyle\int\limits_{L_+}\dfrac{\mu_+(\zeta_+)d\zeta_+} {2(w-\zeta_+)}\\[2ex]
\lambda_-(\eta,w)=\sqrt{\dfrac{w-\eta}{w-\eta_0}},\quad
\lambda_-(\eta,w=\infty) =1,&&
\sigma_-(\eta,w)=\displaystyle\int\limits_{L_-}\dfrac{\mu_-(\zeta_-)d\zeta_-} {2(w-\zeta_-)}
\end{array}
\end{equation}
One can infer from these expressions that the determinant of  $\mathbf{\Psi}(\xi,\eta,w)$ does not vanish anywhere in the region  $\Omega^3\backslash (\mathcal{L}_+\cup \mathcal{L}_-)$ and therefore,  $\mathbf{\Psi}^{-1}(\xi,\eta,w)$ is also holomorphic in this region.

The typical section of the region $\Omega^3\backslash \mathcal{L}$ by the plane $\{\xi=\hbox{const}, \eta=\hbox{const}\}$, called here simply as the spectral plane, is shown on the Figure 1, where the structures of the local regions (\ref{Omegapm}) which contain the local cuts $L_+$ and $L_-$ respectively are shown for the hyperbolic case ($\epsilon=1$) and for the elliptic case ($\epsilon=-1$) separately. For the later convenience, we prescribe to the curves $L_\pm$ the orientation. Namely, $L_+$ starts from $w=\xi_0$ and goes to $w=\xi$, and $L_-$ starts at $w=\eta_0$ and goes to $w=\eta$. It is useful to note that in the hyperbolic case for different space-time regions the relative positions on the real axis of the points $w=\xi_0$ and $w=\xi$  as well as of the points  $w=\eta_0$ and $w=\eta$ can differ from those shown on the figure.

\begin{figure}[h]
\begin{center} \epsfxsize=.8\textwidth \epsfbox{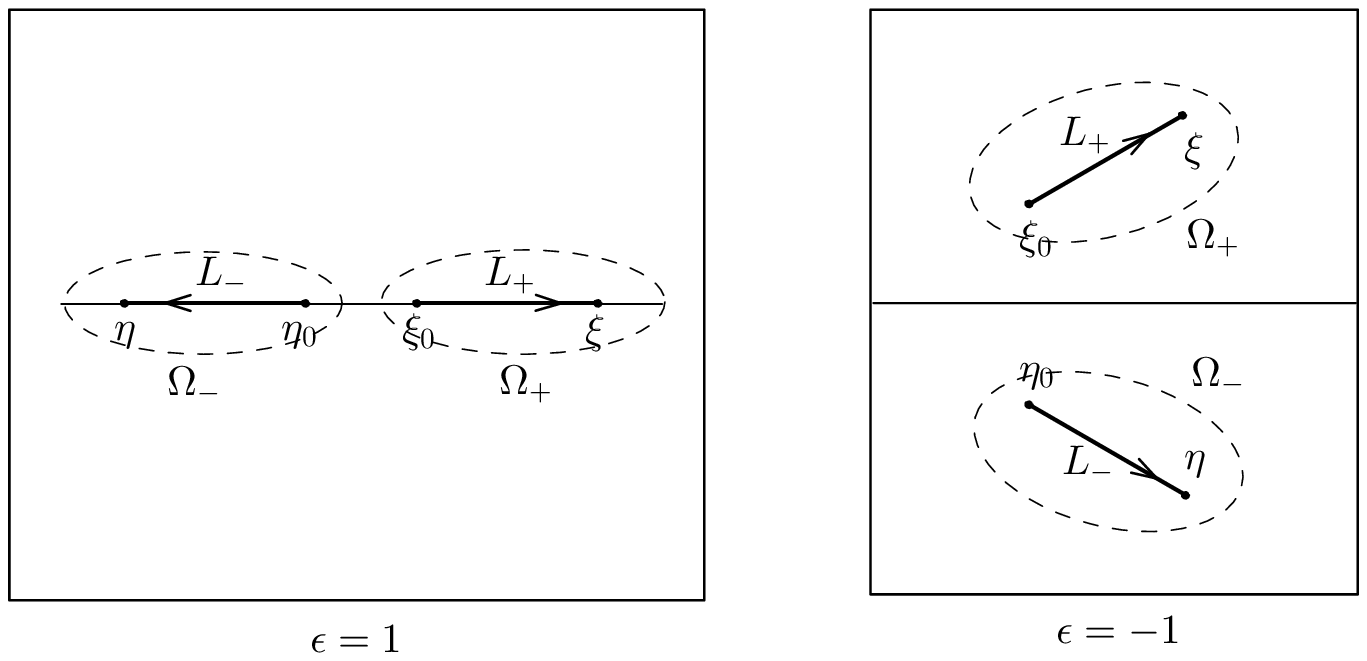}
\end{center}
\caption{ \footnotesize Structure of the compound cut on the spectral plane for the hyperbolic case ($\epsilon=1$) and for the elliptic case ($\epsilon=-1$). The local regions $\Omega_\pm$ are shown by the dash lines which surround the corresponding cuts $L_\pm$. In the hyperbolic case for different space-time regions the order of location on the real axis of the points $w=\xi_0$ and $w=\xi$ and similarly, of the points $w=\eta_0$ and $w=\eta$ can be  opposite.
\label{Ris1}}
\end{figure}
\paragraph{The branching of $\mathbf{\Psi}(\xi,\eta,w)$ and $\mathbf{\Psi}^{-1}(\xi,\eta,w)$ at the ``endpoints'' of the cuts $\mathcal{L}_+$ and $\mathcal{L}_-$.}
As far as the canonical Jordan forms of the matrices $\mathbf{U}$ and $\mathbf{V}$ are determined by the expressions (\ref{CanonicalForms}),  $\xi=w$ and $\eta=w$ are the branching points for  $\mathbf{\Psi}_+(\xi,w)$ and $\mathbf{\Psi}_-(\eta,w)$ respectively and therefore (because of the factorization (\ref{Factorization})), they are the branching points for $\mathbf{\Psi}(\xi,\eta,w)$ as well.
Near these points, i.e. for any $w\in\Omega_+$ and the values of $\xi$ which are close enough to $w$, as well as for any $w\in\Omega_-$ and the values of $\eta$ which are close enough to these $w$, the non-normalized yet fundamental solutions of (\ref{LinSys}) can be represented respectively by some locally convergent series
\begin{equation}\label{LocalSeriesA}
\begin{array}{l}\mathbf{\Psi}(\xi,\eta,w)=(\xi-w)^{r_+} \left[a_0 +a_1(\xi-w)+a_2(\xi-w)^2+\ldots\right]\\[1ex]
\phantom{\mathbf{\Psi}(\xi,\eta,w)=}+ b_0+b_1(\xi-w)+b_2(\xi-w)^2+\ldots\\[0ex]
\mathbf{\Psi}(\xi,\eta,w)=(\eta-w)^{r_-} \left[c_0 +c_1(\eta-w)+c_2(\eta-w)^2+\ldots\right]\\[1ex]
\phantom{\mathbf{\Psi}(\xi,\eta,w)=}+ d_0+d_1(\eta-w)+d_2(\eta-w)^2+\ldots
\end{array}
\end{equation}
where all coefficients of the first expansion are holomorphic functions of the parameters $\eta$ and $w$, and of the second expansion -- of the parameters $\xi$ and $w$, because these coefficients can be expressed algebraically in terms of the values $\mathbf{U}(\xi=w,\eta)$ and $\mathbf{V}(\xi,\eta=w)$ respectively and their derivatives with respect to $w$. The values $r_+$  and $r_-$ of the exponents in these expressions are determined by the eugenvalues of the matrices  $\mathbf{U}$ and $\mathbf{V}$ (with the multipliers  $2i$ in the left hand sides of the equations (\ref{LinSys}) taken into account) and both of them occur to be equal to $-\frac{1}{2}$ in the absence of the spinor field ($\mu_+(\xi)\equiv 0$, $\mu_-(\eta)\equiv 0$) or $-\frac{1}{2}+\frac{i}{2}\mu_+(w)$ and $-\frac{1}{2}+\frac{i}{2}\mu_-(w)$ respectively in the case of its presence. It is easy to show that for  $\mathbf{\Psi}^{-1}(\xi,\eta,w)$ we have the expansions  (\ref{LocalSeriesA}), in which, however, it is necessary to substitute  $r_+\to-r_+$ and $r_-\to-r_-$.

A restriction of the domains $\Omega_{\xi_0}$ and $\Omega_{\eta_0}$ of the variables $\xi$ and $\eta$ by some small enough neighborhoods of their initial values $\xi_0$ and $\eta_0$ respectively provides the points $\xi=\xi_0$ and $\eta=\eta_0$ to be located inside the circles of convergence of the series (\ref{LocalSeriesA}) for $\mathbf{\Psi}(\xi,\eta,w)$ and of similar expansions for $\mathbf{\Psi}^{-1}(\xi,\eta,w)$.
Because in this case the whole cut $\mathcal{L}_+$ is inside of the circle of convergence of the first expansion in (\ref{LocalSeriesA}) and the whole cut $\mathcal{L}_-$ is inside of the circle of convergence of the second of these expansions, we can use these expansions in the expression (\ref{NormPsi}) and calculate the corresponding expansion for the normalized fundamental solution. If we take into account also that the normalized solution should be holomorphic in $\Omega^3$ outside the cuts $\mathcal{L}_+$ and $\mathcal{L}_-$, we obtain that in the neighborhood of   $\mathcal{L}_\pm$ the normalized fundamental solution $\mathbf{\Psi}(\xi,\eta,w)$ and $\mathbf{\Psi}^{-1}(\xi,\eta,w)$ take the forms:
\begin{equation}\label{LocalA}
\begin{array}{lll}
\mathcal{L_+:}&
\mathbf{\Psi}(\xi,\eta,w)=A \left(\dfrac{\xi-w}{\xi_0-w} \right)^{r_+} \hskip-1ex + B,&
\mathbf{\Psi}^{-1}(\xi,\eta,w)=\widetilde{A}\left(\dfrac{\xi-w}{\xi_0-w}
\right)^{-r_+}\hskip-1ex  +\widetilde{B}\\[2ex]
\mathcal{L_-:}&
\mathbf{\Psi}(\xi,\eta,w)=C \left(\dfrac{\eta-w}{\eta_0-w} \right)^{r_-}\hskip-1ex  + D,&
\mathbf{\Psi}^{-1}(\xi,\eta,w)=\widetilde{C}\left(\dfrac{\eta-w}{\eta_0-w}
\right)^{-r_-}\hskip-1ex  +\widetilde{D}
\end{array}
\end{equation}
where the functions $A(\xi,\eta,w)$, $B(\xi,\eta,w)$, $\widetilde{A}(\xi,\eta,w)$  and $\widetilde{B}(\xi,\eta,w)$  are holomorphic in the neighborhood of the cut $\mathcal{L}_+$, and the functions $C(\xi,\eta,w)$, $D(\xi,\eta,w)$, $\widetilde{C}(\xi,\eta,w)$  and $\widetilde{D}(\xi,\eta,w)$ -- in the neighborhood of the cut $\mathcal{L}_-$. We note that the expressions for the determinants of $\mathbf{\Psi}$ derived above suggest useful modifications of the expressions (\ref{LocalA}). Namely, if we redefine the holomorphic coefficients in the expressions (\ref{LocalA}), the branching multipliers in these expressions can be substituted by the new ones which can be expressed in terms of the functions $\lambda_\pm$ and $\sigma_\pm$ defined in (\ref{Lambdasigma}):
\begin{equation}\label{LocalB}
\begin{array}{lclcclccl}
\mathcal{L_+:}&&
\mathbf{\Psi}=\lambda_+^{-1} e^{-i\sigma_+} A + B&&&\mathcal{L_-:}&&\mathbf{\Psi}=\lambda_-^{-1} e^{-i\sigma_-} C + D\\[2ex]
&&\mathbf{\Psi}^{-1}=\lambda_+ e^{i\sigma_+}\widetilde{A}+\widetilde{B}
&&& &&\mathbf{\Psi}^{-1}=\lambda_- e^{i\sigma_-}\widetilde{C}+\widetilde{D} \end{array}
\end{equation}
Later we shall elaborate the structures of the coefficients in these local representations, however, even now it can be seen from (\ref{LocalB}) that on the spectral plane, for given $\xi$ and $\eta$, the functions $\mathbf{\Psi}$ and $\mathbf{\Psi}^{-1}$ have finite left and right limits at the internal points of the contours $L_+$ and $L_-$. The half of the differences of these limits (i.e. the jumps of these functions on these contours) satisfy the H\"older condition.
In the absence of the spinor fields we have $\mu_+(\xi)\equiv 0$, $\mu_-(\eta)\equiv 0$ and  $r_+=r_-=-1/2$. Therefore at the initial points $w=\xi_0$ and $w=\eta_0$ of the contours $L_+$ and $L_-$ the jumps of $\mathbf{\Psi}(\xi,\eta,w)$ vanish and they have the integrable infinities at the endpoints $w=\xi$ and $w=\eta$. The jumps of the function $\mathbf{\Psi}^{-1}(\xi,\eta,w)$ are of just opposite character: they have the integrable infinities at the initial points   $w=\xi_0$ and $w=\eta_0$ of the contours $L_+$ and $L_-$ and they vanish at the endpoints $w=\xi$ and $w=\eta$ of these contours.
In the presence of the spinor field, its inputs into the values of $r_+$ and $r_-$ in the hyperbolic case are pure imaginary and in this case the behaviour of the jumps of the functions  $\mathbf{\Psi}$ and $\mathbf{\Psi}^{-1}$ at the endpoints of the contours is qualitatively similar to the previous case. However, in the elliptic case this is not the case, and it is necessary to impose some restrictions on the behaviour of the spinor field for the conditions of integrability of the jumps $\vert\hbox{Re}\,r_+\vert \le 1$ and $\vert\hbox{Re}\,r_-\vert \le 1$ to be satisfied. The described above properties of $\mathbf{\Psi}$ and $\mathbf{\Psi}^{-1}$ in the region $\Omega^3$ can be summarized in a theorem which leads to some obvious but important  consequences. \\

$\underline{\hbox{\textsc{Theorem 1.}}}$\quad
\textit{For any local solution $\mathbf{U}(\xi,\eta)$, $\mathbf{V}(\xi,\eta)$ of the ``null-curvature'' equations  (\ref{NullCurv}) with the Jordan conditions (\ref{CanonicalForms}) there exists a neighborhood  $\Omega^2=\Omega_{\xi_0}\times\Omega_{\eta_0}$
of the initial point $P_0$ in $\mathbb{C}^2$, where the fundamental solution $\mathbf{\Psi}(\xi,\eta,w)$ of the linear system (\ref{LinSys}) with a condition (\ref{CanonicalForms}) exists, is unique and
\begin{itemize}
\item $\mathbf{\Psi}(\xi,\eta,w)$ is holomorphic in the region $(\Omega^2\times\overline{\mathbb C})\backslash ({\cal L}_+\cup {\cal L}_-)$,
\item $\mathbf{\Psi}(\xi_0,\eta_0,w)=\mathbf{I}$\quad and \quad
$\mathbf{\Psi}(\xi,\eta,w=\infty)=\text{\bf I}$,
\item{$\mathbf{\Psi}(\xi,\eta,w)$ has finite left and right limits on the cuts $L_+$ and $L_-$. The difference of these limits satisfies the H\"older condition and is integrable near the endpoints. (In the elliptic case the last statement holds if the spinor field satisfies  $\vert\hbox{Im}\,\mu_+(\xi)\vert <1$ or equivalently,  $\vert\hbox{Im}\,\mu_-(\eta)\vert< 1$)}
\item{The function $\mathbf{\Psi}^{-1}(\xi,\eta,w)$ also has all these properties.}
\end{itemize}
}
\smallskip

$\underline{\hbox{\textsc{Corollary 1.}}}$\quad
\textit{In the region $(\Omega^2\times\overline{\mathbb C})\backslash ({\cal L}_+\cup {\cal L}_-)$ the normalized fundamental solution  $\mathbf{\Psi}(\xi,\eta,w)$ of linear system (\ref{LinSys}) with conditions (\ref{CanonicalForms}), as well as the inverse matrix  $\mathbf{\Psi}^{-1}(\xi,\eta,w)$ can be expressed as the Cauchy type integrals of their jumps over the cut $L=L_+\cup L_-$ on the spectral plane:
\begin{equation}\label{PsiInts}
\mathbf{\Psi}=\mathbf{I}+\dfrac 1{i\pi}
\int\limits_L\dfrac {[\mathbf{\Psi}]_\zeta}{\zeta-w}d\zeta,\qquad
\mathbf{\Psi}^{-1}=\mathbf{I}+\dfrac 1{i\pi}
\int\limits_L\dfrac {[\mathbf{\Psi}^{-1}]_\zeta}{\zeta-w}d\zeta
\end{equation}}
Here $w$ is outside $L$, the variable $\zeta$ runs over the cut  $L$ which consists of two nonintersecting curves $L_+$ and $L_-$, and $[\ldots]_\zeta$ means the jump of a function, i.e. a half of the difference of its left and right limits at the point  $\zeta\in L$.

\subsection*{Local structure of $\mathbf{\Psi}(\xi,\eta,w)$ near the cuts $\mathcal{L}_+$ and $\mathcal{L}_-$}

A direct consequence of the universal structure (\ref{CanonicalForms}) of the canonical Jordan forms of $\mathbf{U}(\xi,\eta)$ and $\mathbf{V}(\xi,\eta)$ is that among the linear independent solutions whose linear combinations represent the columns of the normalized fundamental solution $\mathbf{\Psi}(\xi,\eta,w)$, there exists only one solution for $N=2,3$ or there exist $d$ solutions for $N=2 d$ which are branching near the points $\xi=w$, while the others are not branching at these points. The same number of branching solutions exists near the points $\eta=w$. We construct now two locally defined auxiliary fundamental solutions $\mathbf{\Psi}_{(+)}$ and $\mathbf{\Psi}_{(-)}$. For the first of them, defined in the neighborhood of $\xi=w$, in the case $N=2,3$ we put  the branching solution in the first column, while the other columns will consist of the non-branching solutions only. For the case $N=2 d$ we put $d$ branching solutions at the first $d$ columns and the last $d$ columns of this fundamental solution will consist of non-branching solutions only. The other fundamental solution defined in the neighborhood of $\eta=w$ is constructed similarly to the first one using the branching and non-branching solutions at these points. It is easy to see that a similar splitting of branching and non-branching solutions will take place between the rows of the matrices which are inverse to these fundamental solutions.  The structure of $\mathbf{\Psi}_{(+)}$ near $\xi=w$ can be shown for $N=3$ by the following expressions (the case $N=2$ arises, if we omit the third rows and columns of all matrices):
$$\begin{array}{l}
\mathbf{\Psi}_{(+)}=\dfrac 1{\Sigma_+} \begin{pmatrix}
\ast \\ \ast\\ \ast\end{pmatrix}\otimes (1,0,0)+
\begin{pmatrix}
0&\ast&\ast \\0&\ast&\ast\\ 0&\ast&\ast\end{pmatrix},\\[4ex]
\mathbf{\Psi}_{(+)}^{-1}=\Sigma_+ \begin{pmatrix}
1 \\0\\0\end{pmatrix}\otimes\begin{pmatrix}
\ast & \ast & \ast\end{pmatrix}+
\begin{pmatrix}
0&0&0 \\ \ast&\ast&\ast\\ \ast&\ast&\ast\end{pmatrix},
\end{array}
$$
where $\Sigma_{(+)}=(\xi-w)^{-r_+}$, the symbol $\otimes$ denotes the tensor product and $\ast$ denotes the components holomorphic near $\xi=w$. For $N=2 d$ the structure of these matrices changes as it is shown here for $d=2$
\begin{equation}\label{Psiplus}
\begin{array}{l}
\mathbf{\Psi}_{(+)}=(\xi-w)^{-\frac 12} \begin{pmatrix}
\ast&\ast \\ \ast&\ast\\ \ast&\ast\\ \ast&\ast\end{pmatrix}\otimes
\begin{pmatrix} 1&0&0&0\\ 0&1&0&0\end{pmatrix}+
\begin{pmatrix}
0&0&\ast&\ast \\0&0&\ast&\ast\\ 0&0&\ast&\ast\\ 0&0&\ast&\ast\end{pmatrix},\\[5ex]
\mathbf{\Psi}_{(+)}^{-1}=(\xi-w)^{\frac 12} \begin{pmatrix}
1&0 \\0&1\\0&0\\0&0\end{pmatrix}\otimes\begin{pmatrix}
\ast & \ast & \ast&\ast\\
\ast & \ast & \ast&\ast\end{pmatrix}+
\begin{pmatrix}
0&0&0&0 \\0&0&0&0 \\ \ast&\ast&\ast&\ast\\ \ast&\ast&\ast&\ast\end{pmatrix},
\end{array}
\end{equation}
where $\otimes$ is the usual matrix multiplication or the tensor product symbol, however, in the last case all $d\times 2d$, $2d\times d$ and $2d\times 2d$  matrices should be considered as two-dimensional, but having the $d\times d$-matrix valued components.
Another auxiliary fundamental solution $\mathbf{\Psi}_{(-)}$ possesses the similar structure near $\eta=w$. These local structures can be expressed in the form
\begin{equation}\label{Psipm}
\begin{array}{lll}
\mathcal{L}_+:&\mathbf{\Psi}_{(+)}=\dfrac 1{\Sigma_+}\,{\bpsi}_{(+)}\otimes \mathbf{k}_{(+)}+\mathbf{M}_{(+)},&
\mathbf{\Psi}^{-1}_{(+)}=\Sigma_+\,\mathbf{l}_{(+)}\otimes
\bphi_{(+)} +\mathbf{N}_{(+)}\\[1ex]
\mathcal{L}_-:&
\mathbf{\Psi}_{(-)}= \dfrac 1{\Sigma_-}\,{\bpsi}_{(-)}\otimes \mathbf{k}_{(-)}+\mathbf{M}_{(-)},
&\mathbf{\Psi}^{-1}_{(-)}= \Sigma_-\,\mathbf{l}_{(-)}\otimes\bphi_{(-)} +\mathbf{N}_{(-)}\end{array}
\end{equation}
where the components of the columns (or $2d\times d$-matrices) $\bpsi_{(\pm)}$, the rows (or $d\times 2d$-matrices) $\bphi_{(\pm)}$ and the matrices $\mathbf{M}_{\pm}$ and $\mathbf{N}_{\pm}$ are holomorphic on the corresponding cuts, and their structure (the location of zeros and units) can be read off from (\ref{Psiplus}) or from the similar representations for $\mathbf{\Psi}_{(-)}$ and $\mathbf{\Psi}^{-1}_{(-)}$.
This allows us to observe that the fragments of the algebraic structures of the auxiliary fundamental solutions introduced in (\ref{Psipm}) should satisfy on the corresponding cuts the following algebraic relations:
\begin{equation}\label{FragPlus}
\begin{array}{lclcl}
\mathcal{L}_+:&&\mathbf{M}_{(+)}\cdot \mathbf{l}_{(+)}=0,&&
\mathbf{k}_{(+)}\cdot \mathbf{N}_{(+)}=0,\\
&&\bphi_{(+)}\cdot\mathbf{M}_{(+)}=0,&&
\mathbf{N}_{(+)}\cdot \bpsi_{(+)}=0,
\end{array}
\end{equation}
and the similar relations should be satisfied on another cut:
\begin{equation}\label{FragMinus}
\begin{array}{lclclclcl}
\mathcal{L}_-:&&\mathbf{M}_{(-)}\cdot \mathbf{l}_{(-)}=0,&&
\mathbf{k}_{(-)}\cdot \mathbf{N}_{(-)}=0,\\
&&\bphi_{(-)}\cdot\mathbf{M}_{(-)}=0,&&
\mathbf{N}_{(-)}\cdot \bpsi_{(-)}=0.
\end{array}
\end{equation}

Thus, in the neighborhood of each of the cuts we have two fundamental solutions. One of them is the ``globally'' defined normalized fundamental solution $\mathbf{\Psi}$. Another one is the locally defined fundamental solution $\mathbf{\Psi}_{(+)}$, if we consider the cut $\mathcal{L}_+$, or $\mathbf{\Psi}_{(-)}$, if we consider the cut  $\mathcal{L}_-$. Any two fundamental solutions in the intersection of their domains are related by some linear transformation, which can depend on the spectral parameter:
$$
\mathbf{\Psi}(\xi,\eta,w)=
\left\{\begin{array}{l}
\mathbf{\Psi}_{(+)}(\xi,\eta,w)\mathbf{C}_+(w),
\\[1ex]
\mathbf{\Psi}_{(-)}
(\xi,\eta,w) \mathbf{C}_-(w),
\end{array}\right.
$$
It is clear that the matrices $\mathbf{C}_\pm$, which normalize the solutions $\mathbf{\Psi}_{(+)}$ and $\mathbf{\Psi}_{(-)}$ should possess the expressions
$\mathbf{C}_+(w)=\mathbf{\Psi}_{(+)}^{-1}(\xi_0,\eta_0,w)$ and
$\mathbf{C}_-(w)=\mathbf{\Psi}_{(-)}^{-1}(\xi_0,\eta_0,w)$.
These expressions, together with (\ref{Psipm}), allow to obtain the similar local representations for $\mathbf{\Psi}$ and $\mathbf{\Psi}^{-1}$ as well as to show that the corresponding fragments of the local structures of these matrix functions satisfy the relations which are similar to  (\ref{FragPlus}) and (\ref{FragMinus}). Let us formulate now the properties of $\mathbf{\Psi}$ and $\mathbf{\Psi}^{-1}$ described above as the following theorem, which consequences lead to explicit expressions for the jumps of these functions on the cuts $\mathcal{L}_+$ and $\mathcal{L}_-$.\\

$\underline{\hbox{\textsc{Theorem 2.}}}$\quad
\textit{For any local solution $\mathbf{U}$, $\mathbf{V}$ of the ``null-curvature'' equations (\ref{NullCurv}) with the condition  (\ref{CanonicalForms}), there exists a neighborhood $\Omega^2=\Omega_{\xi_0}\times\Omega_{\eta_0}$ of the initial point $P_0$, such that
\begin{itemize}
\item the normalized fundamental solution  $\mathbf{\Psi}(\xi,\eta,w)$ of the system (\ref{LinSys}) in the regions $\Omega^2\times\Omega_+$ and $\Omega^2\times\Omega_-$ (the neibourhoods of the cuts $\mathcal{L}_+$ and $\mathcal{L}_-$) has the local structure
\begin{equation}\label{LocStrucA}
\mathbf{\Psi}=\left\{\begin{array}{ll}
\lambda_+^{-1} e^{-i\sigma_+} \bpsi_+(\xi,\eta,w)\otimes \mathbf{k}_+(w)+\mathbf{M}_+(\xi,\eta,w),& w\in\Omega_+\\[1ex]
\lambda_-^{-1} e^{-i\sigma_-} \bpsi_-(\xi,\eta,w)\otimes \mathbf{k}_-(w)+\mathbf{M}_-(\xi,\eta,w),& w\in\Omega_-
\end{array}\right.
\end{equation}
and the function $\mathbf{\Psi}^{-1}(\xi,\eta,w)$ has the similar local structure:
\begin{equation}\label{LocStrucB}
\mathbf{\Psi}^{-1}=\left\{\begin{array}{ll}
\lambda_+ e^{i\sigma_+} \mathbf{l}_+(w)\otimes \bphi_+(\xi,\eta,w) +\mathbf{N}_+(\xi,\eta,w),&w\in\Omega_+\\[1ex]
\lambda_- e^{i\sigma_-} \mathbf{l}_-(w)\otimes \bphi_-(\xi,\eta,w) +\mathbf{N}_-(\xi,\eta,w),& w\in\Omega_-
\end{array}\right.
\end{equation}
where the row-vectors (for $N=2,3$) or $d\times 2d$-matrices (for $N=2d$) $\mathbf{k}_\pm(w)$, $\bphi_\pm(\xi,\eta,w)$, the column-vectors (for $N=2,3$) or $2d\times d$-matrices (for $N=2d$) $\mathbf{l}_\pm(w)$, $\bpsi_\pm(\xi,\eta,w)$, as well as the matrices $\mathbf{M}_\pm(\xi,\eta,w)$,  $\mathbf{N}_\pm(\xi,\eta,w)$ are holomorphic in the neighborhoods of $\mathcal{L}_\pm$ respectively, and the functions $\lambda_\pm$ and $\sigma_\pm$ have been defined in (\ref{Lambdasigma});
the symbol $\otimes$ for $N=2,3$ means the tensor multiplication but for $N=2d$ it should be understood as the usual matrix multiplication;
\item in the neighborhoods of $\mathcal{L}_+$ and $\mathcal{L}_-$ the following relations are satisfied:
$$
\hskip-2ex\begin{array}{lcllll}
\Omega^2\times\Omega_+:&&
\mathbf{M}_+\cdot\mathbf{l}_+=0,& \mathbf{k}_+\cdot\mathbf{N}_+=0,&
\bphi_+\cdot\mathbf{M}_+=0,&
\mathbf{N}_+\cdot\bpsi_+=0,\\[1ex]
\Omega^2\times\Omega_-:&&
\mathbf{M}_-\cdot\mathbf{l}_-=0,& \mathbf{k}_-\cdot\mathbf{N}_-=0,&
\bphi_-\cdot\mathbf{M}_-=0,&
\mathbf{N}_-\cdot\bpsi_-=0;
\end{array}
$$
\item at the initial point $P_0(\xi_0,\eta_0)$ the values of the functions  $\bphi_\pm$, $\bpsi_\pm$, $\mathbf{M}_\pm$, $\mathbf{N}_\pm$
can be expressed in terms of the functions  $\mathbf{k}_\pm$, $\mathbf{l}_\pm$:
\begin{equation}\label{InitialValues}
\begin{array}{lcl}
\bpsi_{0\pm}=\mathbf{l}_\pm(w)\cdot(\mathbf{k}_\pm(w)\cdot \mathbf{l}_\pm(w))^{-1},&&\mathbf{M}_{0\pm}=\mathbf{I}-\bpsi_{0\pm} \otimes\mathbf{k}_{\pm}(w),\\[1ex]
\bphi_{0\pm}=(\mathbf{k}_\pm(w)\cdot \mathbf{l}_\pm(w))^{-1}\cdot\mathbf{k}_\pm(w),&&
\mathbf{N}_{0\pm}=\mathbf{I}-\mathbf{l}_{\pm}(w)\otimes \bphi_{0\pm},
\end{array}
\end{equation}
where the upper (or lower) signs must be chosen simultaneously in both sides of all the equations.
\end{itemize}
}

\noindent
For a function with a jump on the cut $L$, we let the brackets and braces denote
the respective values of its jump and ``continious part''  at the point $\tau\in L$. These values are respectively defined as a half the difference and the half  the sum of the left and right limits of this function at the point of the cut. From the definitions (\ref{Lambdasigma}) we then obtain:
$$
\begin{array}{llll}
\{\lambda_+\}_{\tau_+}=0,&\{\lambda_+^{-1}\}_{\tau_+}=0, &[\sigma_+]_{\tau_+}=-i\dfrac \pi 2 \mu_+(\tau_+),&\{\sigma_+\}_{\tau_+}=-\dfrac 12\vpint\limits_{L_+}\dfrac{\mu_+(\zeta_+)d\zeta_+} {\zeta_+-\tau_+}\\[1ex]
\{\lambda_-\}_{\tau_-}=0,&\{\lambda_-^{-1}\}_{\tau_-}=0, &[\sigma_-]_{\tau_-}=-i\dfrac \pi 2 \mu_-(\tau_-),&\{\sigma_-\}_{\tau_-}=-\dfrac 12\vpint\limits_{L_-}\dfrac{\mu_-(\zeta_-)d\zeta_-} {\zeta_--\tau_-}
\end{array}
$$
where $\tau_+\in L_+$, $\tau_-\in L_-$, and the integrals here are the Cauchy principle value integrals. For the jumps and continuous parts of the products of these functions at the points $\tau_+\in L_+$ and $\tau_-\in L_-$ it is convenient to use more explicit expressions of the form:
$$
\begin{array}{lcl}
[\lambda e^{i\sigma}]=[\lambda] e^{i\{\sigma\}}\cos[\sigma]&&
[\lambda^{-1} e^{-i\sigma}]=[\lambda^{-1}] e^{-i\{\sigma\}}\cos[\sigma]\\[1ex]
\{\lambda e^{i\sigma}\}=i[\lambda] e^{i\{\sigma\}}\sin[\sigma]&&
\{\lambda^{-1} e^{-i\sigma}\}=-i[\lambda^{-1}] e^{-i\{\sigma\}}\sin[\sigma]
\end{array}
$$
Using these relations we obtain from the Theorem 2 the following\\[2ex]
\medskip
$\underline{\hbox{\textsc{Corollary 2.}}}$\quad
\textit{The jumps of  $\mathbf{\Psi}$ and $\mathbf{\Psi}^{-1}$ on  $L_\pm$ have the expressions
$$
[\mathbf{\Psi}]_{L_\pm}=
[\lambda_\pm^{-1}] e^{-i\{\sigma_\pm\}}\cos [\sigma_\pm]
\bpsi_\pm\otimes\mathbf{k}_\pm, \qquad
[\mathbf{\Psi}]_{L_\pm}^{-1}=[\lambda_\pm] e^{i\{\sigma_\pm\}}
\cos [\sigma_\pm]
\mathbf{l}_\pm\otimes\bphi_\pm,
$$
and the similar expressions for the continuous parts of these functions yield the representations for the matrix functions  $\mathbf{M}_\pm$ and $\mathbf{N}_\pm$ at the points of the cuts  $\mathcal{L}_\pm$:
$$
\begin{array}{lll}
\mathbf{M}_\pm&=&\{\mathbf{\Psi}\}_{L_\pm}
+i[\lambda_\pm^{-1}] e^{-i\{\sigma_\pm\}}\sin [\sigma_\pm]
\,\,\bpsi_\pm\otimes\mathbf{k}_\pm,\\[0ex]
\mathbf{N}_\pm&=&\{\mathbf{\Psi}^{-1}\}_{L_\pm}-
i[\lambda_\pm] e^{i\{\sigma_\pm\}}\sin [\sigma_\pm]
\,\,\mathbf{l}_\pm\otimes\bphi_\pm.
\end{array}
$$
where the upper (or lower) signs must be chosen simultaneously in both sides, and the jumps and the continuous parts of  $\lambda_+$, $\sigma_+$ and $\lambda_-$, $\sigma_-$ must be calculated on the respective cuts $L_+$ and $L_-$.}\\

\subsection*{``Coordinates'' in the spaces of local solutions}
\paragraph{Independent functional parameters.} Among the fragments of the local structures of $\mathbf{\Psi}$  and $\mathbf{\Psi}^{-1}$, there are four vector (for $N=2,3$) or $d\times 2d$ and $2d\times d$ - matrix (for $N=2 d$) functions depending on the spectral parameter only. These are the functions $\mathbf{k}_+(w)$, $\mathbf{l}_+(w)$ holomorphic in the region $\Omega_+$ and the functions $\mathbf{k}_-(w)$, $\mathbf{l}_-(w)$   holomorphic in the region $\Omega_-$. It is easy to see from  (\ref{LocStrucA}), (\ref{LocStrucB}) that these functions are defined in a ``projective'' sense only, i.e. in the vector case   ($N=2,3$) each of these functions is defined up to its multiplication by an arbitrary scalar function of $w$, which is holomorphic in the neighborhood of the corresponding cut, and  division of another multiplier by the same function. In the matrix case ($N=2d$) this multiplier is an arbitrary holomorphic  nondegenerate $d\times d$ - matrix function of $w$. This arbitrariness can be eliminated if we put  for the case $N=2,3$ the first components of the vectors $\mathbf{k}_\pm(w)$ and $\mathbf{l}_\pm(w)$ equal to $1$, and for $N=2d$ the first $d$-dimensional blocks of these matrices equal to a unit matrix:
\begin{equation}\label{ParametrizationA}
\begin{array}{lcl}
\mathbf{k}_\pm^{N=2}(w)=\{1,\mathbf{u}_\pm(w)\},&&
\mathbf{l}_\pm^{N=2}(w)=\{1,\mathbf{p}_\pm(w)\}^T,\\[1ex]
\mathbf{k}_\pm^{N=3}(w)=\{1,\mathbf{u}_\pm(w),\mathbf{v}_\pm(w)\},&&
\mathbf{l}_\pm^{N=3}(w)=\{1,\mathbf{p}_\pm(w),\mathbf{q}_\pm(w)\}^T,\\[1ex]
\mathbf{k}_\pm^{N=2d}(w)=\{I_d,\,\,\mathbf{u}_\pm(w)\},&&
\mathbf{l}_\pm^{N=2d}(w)=\{I_d,\,\,\mathbf{p}_\pm(w)\}^T,
\end{array}
\end{equation}
where in the last case $\mathbf{u}_\pm$ and $\mathbf{p}_\pm$ are  $d\times d$-matrix functions and ${}^T$ denotes the matrix transposition. The components of $\mathbf{k}_\pm(w)$ and $\mathbf{l}_\pm(w)$, which can be supplemented for $N=2,3$ by  $\mu_\pm(w)$, constitute, as we shall see below, a complete set of independent functional parameters which can serve as the ``coodinates'' in the space of local solutions of the ``null-curvature'' equations (\ref{NullCurv}) with the conditions (\ref{CanonicalForms}).

\paragraph{The monodromy data as functional parameters.}
In view of the expressions (\ref{InitialValues}), the vector (or matrix) functions $\mathbf{k}_\pm(w)$ and $\mathbf{l}_\pm(w)$ admit a simple interpretation as the initial values (defined in the ``projective'' sense mentioned above) for the branching parts of the linearly independent solutions which constitute the normalized fundamental solution  $\mathbf{\Psi}$ and $\mathbf{\Psi}^{-1}$. The independent components of these functions admit also another, more clear interpretation as a complete set of data which determine the monodromy of the normalized fundamental solution $\mathbf{\Psi}$ on the spectral plane. To show this we consider on the spectral plane  two paths $t_+$ and $t_-$, which go, say for definiteness, in the clockwise direction from the left edge of the cut $L_+$ or $L_-$ to its right edge around the points $w=\xi$ and $w=\eta$ respectively. After the analytic continuation along each of these paths of the normalized fundamental solution $\mathbf{\Psi}$ of the system (\ref{LinSys}), we obtain another fundamental solutions  $\widetilde{\mathbf{\Psi}}$, which are connected with $\mathbf{\Psi}$ by some linear transformations  $\mathbf{\Psi}\stackrel {t_\pm}\longrightarrow \widetilde{\mathbf{\Psi}}=\mathbf{\Psi}\cdot \mathbf{T}_\pm(w)$. The matrices of these transformations (the monodromy matrices) can be calculated using the relation $\mathbf{T}_\pm(w) = \mathbf{\Psi}^{-1}\cdot\widetilde{\mathbf{\Psi}}$, where the local representations (\ref{LocStrucA}), (\ref{LocStrucB}) for $\mathbf{\Psi}$ and $\mathbf{\Psi}^{-1}$ near $L_\pm$ are used. The result of these calculations is
\begin{equation}\label{TMatrices}
\begin{array}{l}
\mathbf{T_+}(w)= \mathbf{I}-(1+e^{-\pi \mu_+(w)})\,\,\mathbf{l}_+(w)\otimes(\mathbf{k}_+(w)\cdot \mathbf{l}_+(w))^{-1}\otimes\mathbf{k}_+(w),\\
\mathbf{T_-}(w)= \mathbf{I}-(1+e^{-\pi \mu_-(w)})\,\,\mathbf{l}_-(w)\otimes(\mathbf{k}_-(w)\cdot \mathbf{l}_-(w))^{-1}\otimes\mathbf{k}_-(w).
\end{array}
\end{equation}
Thus, the monodromy matrices  $\mathbf{T}_\pm(w)$  are expressed in terms of the components of the vectors (matrices) $\mathbf{k}_\pm(w)$, $\mathbf{l}_\pm(w)$ and functions $\mu_\pm(w)$ which determine the spinor field. The inverse statement is also true: all independent functional parameters in the components (\ref{ParametrizationA}) of the vectors $\mathbf{k}_\pm(w)$, $\mathbf{l}_\pm(w)$ and the functions  $\mu_+(w)$ and $\mu_-(w)$ can be determined unambiguously from the components of the monodromy matrices (\ref{TMatrices}). As it follows from the Theorems 1 and 2,  the set of functional parameters  (\ref{ParametrizationA}) together with the functions $\mu_+(w)$ and $\mu_-(w)$, which we call below as the ``extended monodromy data'',   can be determined unambiguously for any local solution of the ``null-curvature'' equations (\ref{NullCurv}) with the conditions (\ref{CanonicalForms}).

The extended monodromy data can play the role which is similar to the role of the scattering data in the inverse scattering method. This allows, similarly to the known ``scattering transform'',  to call the choice of such ``coordinates'' in the space of solutions as the ``monodromy transform''.

\subsection*{Inverse problem of the monodromy transform} We show now that for any choice of the functions of the spectral parameter(\ref{ParametrizationA}),  holomorphic in some neighborhoods of the points  $w=\xi_0$ (for the functions with the suffix ``$+$'') and $w=\eta_0$ (for the functions with the suffix ``$-$'') on the spectral plane, there always exists a unique local solution of the ``null-curvature'' equations (\ref{NullCurv}) with conditions (\ref{CanonicalForms}) for which the normalized fundamental solution of the linear system (\ref{LinSys}) is characterized by these extended monodromy data.

\paragraph{The basic system of linear singular integral equations.}
Let us improve a bit our notations. We supply the parameters
$\tau$, $\zeta$, ranging the curves $L_+$ and $L_-$ with the corresponding suffices $+$ or $-$ and introduce the corresponding functions without suffices for the pairs of functions determined on $L_\pm$ differing by the  suffices $+$ and $-$. For example,
\begin{equation}\label{Doublefunctions}
\mathbf{k}(\tau)=\left\{\begin{array}{ll}
\mathbf{k}_+(\tau),& \tau=\tau_+\in L_+\\
\mathbf{k}_-(\tau),& \tau=\tau_-\in L_-
\end{array}\right.\quad
\bphi(\xi,\eta,\tau)=\left\{\begin{array}{ll}
\bphi_+(\xi,\eta,\tau),& \tau=\tau_+\in L_+\\
\bphi_-(\xi,\eta,\tau),& \tau=\tau_-\in L_-
\end{array}\right.
\end{equation}
The functions $\lambda$, $\sigma$, $\mathbf{l}$, $\bpsi$, $\mathbf{M}$ and $\mathbf{N}$ will be defined analogously.

Using the Sokhotskii - Plemelj formulas, the continuous parts of $\mathbf{\Psi}$ and $\mathbf{\Psi}^{-1}$, represented (in accordance with the Theorem 1 and the Corollary 1) by the integrals (\ref{PsiInts}), can be expressed in the forms
$$
\{\mathbf{\Psi}\}_\tau=\mathbf{I}+\dfrac 1{i\pi}
\vpint\limits_L\dfrac {[\mathbf{\Psi}]_\zeta}{\zeta-\tau}d\zeta,\qquad
\{\mathbf{\Psi}^{-1}\}_\tau=\mathbf{I}+\dfrac 1{i\pi}
\vpint\limits_L\dfrac {[\mathbf{\Psi}^{-1}]_\zeta}{\zeta-\tau}d\zeta
$$
where the integrals over the cut $L=L_+\cup L_-$ are the Cauchy principal value integrals and the variable $\tau=\tau_+\in L_+$ or $\tau=\tau_-\in L_-$. Substitution of these expressions and the expressions  presented in the Corollary 2 in the algebraic constraints  $\mathbf{M}\cdot\mathbf{l}=0$ and $\mathbf{k}\cdot\mathbf{N}=0$, stated by the Theorem 2, leads to the following statement.\\[2ex]
$\underline{\hbox{\textsc{Theorem 3.}}}$\quad
\textit{For any local solution of the ``null-curvature'' equations (\ref{NullCurv}) with the conditions (\ref{CanonicalForms}), the fragments of the algebraic structures of  $\mathbf{\Psi}$ and $\mathbf{\Psi}^{-1}$ on the cuts $L_+$ and $L_-$ must satisfy the integral equations
\begin{equation}\label{IntEqs}
\begin{array}{l}
\nu(\xi,\eta,\tau)\cdot\bphi(\xi,\eta,\tau)+\dfrac 1{\pi i} \vpint\limits_L\dfrac {\mathcal{K}(\xi,\eta,\tau,\zeta)}{\zeta-\tau}\, \cdot\bphi(\xi,\eta,\zeta)\,d\zeta=\mathbf{k}(\tau)\\[2ex]
\bpsi(\xi,\eta,\tau)\cdot\widetilde{\nu}(\xi,\eta,\tau)+\dfrac 1{\pi i} \vpint\limits_L \bpsi(\xi,\eta,\zeta)\cdot\dfrac {\widetilde{\mathcal{K}}(\xi,\eta,\tau,\zeta)}{\zeta-\tau}\, \,d\zeta=\mathbf{l}(\tau)
\end{array}
\end{equation}
where the integrals over $L=L_+\cup L_-$ are the Cauchy principal value integrals, the dot for the case $N=2 d$ denotes a matrix product, and the scalar (for $N=2,3$) or $d\times d$-matrix  (for $N=2d$) kernels and coefficients are
\begin{equation}\label{Coeff}
\begin{array}{lccl}
\mathcal{K}(\xi,\eta,\tau,\zeta)=-[\lambda e^{i\sigma}]_\zeta \mathcal{H}(\tau,\zeta)&&&\nu(\xi,\eta,\tau)=\{\lambda e^{i\sigma}\}_\tau \mathcal{H}(\tau,\tau)
\\[1ex]
\widetilde{\mathcal{K}}(\xi,\eta,\tau,\zeta)=-[\lambda^{-1} e^{-i\sigma}]_\zeta \mathcal{H}(\zeta,\tau)&&&
\widetilde{\nu}(\xi,\eta,\tau)=\{\lambda^{-1} e^{-i\sigma}\}_\tau \mathcal{H}(\tau,\tau)
\end{array}
\end{equation}
where $\mathcal{H}(x,y)\equiv (\mathbf{k}(x)\cdot \mathbf{l}(y))$, and each of the parameters $\zeta$ and $\tau$ ranges the contour $L=L_+\cup L_-$.
}
\bigskip

We note here that each of the integrals in (\ref{IntEqs}) is a sum of two integrals over the cuts $L_+$ and $L_-$, and each of the integral relations (\ref{IntEqs}) in a more detailed form represents two, not one, integral equations, which arise from the original one after a subsequent substitution there of $\tau=\tau_+$ and  $\tau=\tau_-$. Thus, we obtain two pairs of integral equations with four, not two,  different kernels determined by the functions (or $d\times d$-matrices for $N=2d$) of the form $\mathcal{H}_{\pm\pm}(\tau,\zeta)\equiv\mathcal{H}(\tau_\pm,\zeta_\pm)$.

\paragraph{Existence and uniqueness of solution of the inverse problem.}
As we have shown above, for any local solution of the ``null-curvature'' equations (\ref{NullCurv}) with the conditions  (\ref{CanonicalForms}) the relations (\ref{IntEqs}) should be satisfied. However, (\ref{IntEqs}) can be considered as the linear singular integral equations for the components of $\bphi$ and $\bpsi$ for a given set of extended monodromy data, the components (\ref{ParametrizationA}) of the vectors (matrices) $\mathbf{k}$ and $\mathbf{l}$ and the functions $\mu_+(w)$ and $\mu_-(w)$. In this case, it is easy to see that these equations don't constitute a coupled system and, moreover, each of them decouples into independent equations for each of the components of the unknown vector function in the scalar case or, in the matrix case ($N=2d$), for each of two  $d\times d$ - blocks, which constitute $d\times 2d$ and $2d\times d$ - matrices $\bphi$ and $\bpsi$.

The linear singular integral equations derived here are characterized by a specific structure of the integration path which consists of two nonclosed and nonintersecting curves. Another important restriction imposed by the spectral problem conditions is that the solutions $\bphi$ and $\bpsi$ of these equations should be bounded at each of the four end points of the integration contour. In accordance with the general theory of the linear singular integral equations \cite{Muskhelishvili:1968},\cite{Gakhov:1977}, this implies that the characteristic equations corresponding to each of the described above independent scalar or $d\times d$ - matrix integral equations which arise after a decoupling of the equations of the system (\ref{IntEqs}), have zero valued index\footnote{Calculating this index, one should take into account the singularities at the endpoints of the contour which are due to the presence of the jumps of $\lambda$ in the integrands.}. Therefore, the corresponding complete (i.e. including the regular part) integral equation admits a regularization which leads to an equivalent quasi-Fredholm linear integral equation of the second kind. For a construction of solutions of this equation the usual method of subsequent approximations can be used and the existence and uniqueness of local solutions can be proved using a standard way. Namely, for any choice of the monodromy data we can always chose a sufficiently small neighborhood $\Omega_{\xi_0}\times\Omega_{\eta_0}$ of the initial point $P_0(\xi_0,\eta_0)$ where the corresponding integral operator is contracting and the convergence of the subsequent approximations can be proved using a majorant method. However, this detailed proof is beyond the scope of this paper, and we restrict ourself here to formulating the final result.

$\underline{\hbox{\textsc{Theorem 4.}}}$\quad
\textit{For arbitrarily chosen extended monodromy data consisting of two scalar functions $\mu_+(w)$, $\mu_-(w)$ and two pairs of vector (for $N=2,3$) or of only  two pairs of $d\times 2d$ and $2d\times d$ - matrix (for $N=2d$)\footnote{We recall here that for the matrix case $N=2d$ with $d>1$ we do not consider the spinor fields and therefore, in this case we put $\mu_+=0$ and $\mu_-=0$.} functions $\mathbf{k}_+(w)$, $\mathbf{l}_+(w)$ and  $\mathbf{k}_-(w)$, $\mathbf{l}_-(w)$, holomorphic respectively in some neighborhoods $\Omega_+$ and $\Omega_-$ of the points $w=\xi_0$ and $w=\eta_0$ on the spectral plane, there exists some neighborhood $\Omega^2=\Omega_{\xi_0}\times\Omega_{\eta_0}$ of the initial point $P_0(\xi_0,\eta_0)$, where the solutions $\bphi_\pm(\xi,\eta,w)$ and $\bpsi_\pm(\xi,\eta,w)$ of the equations  (\ref{IntEqs}) with the coefficients defined in (\ref{Coeff}), (\ref{Lambdasigma}) with (\ref{Doublefunctions}) taken into account exist and are unique. The matrix functions
\begin{equation}\label{PsiIntsB}
\begin{array}{l}
\mathbf{\Psi}=\mathbf{I}+\dfrac 1{i\pi}
\displaystyle\int\limits_L\dfrac {[\lambda^{-1} e^{-i\sigma}]_\zeta} {\zeta-w}
\bpsi(\xi,\eta,\zeta)\otimes\mathbf{k}(\zeta)d\zeta,\\[2ex]
\widetilde{\mathbf{\Psi}}=\mathbf{I}+\dfrac 1{i\pi}
\displaystyle\int\limits_L\dfrac {[\lambda e^{i\sigma}]_\zeta} {\zeta-w}
\mathbf{l}(\zeta)\otimes\bphi(\xi,\eta,\zeta)d\zeta
\end{array}
\end{equation}
constructed for these solutions have the following properties:
\begin{itemize}
\item these matrices are inverse one to another: $\widetilde{\mathbf{\Psi}}=\mathbf{\Psi}^{-1}$;
\item $\mathbf{\Psi}$ is a fundamental solution of the system  (\ref{LinSys}) with the conditions (\ref{CanonicalForms}) that satisfies the normalization condition (\ref{NormPsi}).
\end{itemize}
}
\paragraph{Calculating solutions of the ``null-curvature'' equations for given extended monodromy data.} Given set of the extended monodromy data (\ref{ParametrizationA}), we have to solve one of the integral equations (\ref{IntEqs}). Then the corresponding local solution of the ``null-curvature'' equations (\ref{NullCurv}) with the conditions (\ref{CanonicalForms}) can be calculated, if we realize, using (\ref{LinSys}), that this solution can be related to the expansion  $\mathbf{\Psi}=\mathbf{I}+\dfrac 1w \mathbf{R}+\ldots$ for $w\to\infty$ by the expressions
\begin{equation}\label{UVR}
\mathbf{U}=2i \partial_\xi \mathbf{R},\quad \mathbf{V}=2i \partial_\eta \mathbf{R},
\end{equation}
where the matrix function $\mathbf{R}(\xi,\eta)$ can be expressed from (\ref{PsiIntsB}) in the form
\begin{equation}\label{Rmatrix}
\mathbf{R}=-\dfrac 1{i\pi}
\displaystyle\int\limits_L [\lambda^{-1} e^{-i\sigma}]_\zeta
\bpsi(\xi,\eta,\zeta)\otimes\mathbf{k}(\zeta)d\zeta=
\dfrac 1{i\pi}
\displaystyle\int\limits_L [\lambda e^{i\sigma}]_\zeta
\mathbf{l}(\zeta)\otimes\bphi(\xi,\eta,\zeta)d\zeta
\end{equation}
and therefore, this solution can be expressed in quadratures in terms of the extended monodromy data and the corresponding solution of one of the integral equations (\ref{IntEqs}).

\subsection*{Spaces of local solutions of integrable reductions of Einstein's field equations}
The above construction allowed expressing the general local solution of the ``null-curvature'' equations (\ref{NullCurv}) with the Jordan conditions (\ref{CanonicalForms}) in quadratures in terms of the extended monodromy data and the corresponding solution of the derived above linear system of singular integral equations. Now it is not difficult to show how this construction should be modified to obtain a similar representation of the general local solution for each of the considered here integrable reductions of Einstein's field equations.

\paragraph{The conditions of existence of the matrix integrals $\mathbf{K}(w)$ and $\mathbf{L}(w)$.} The additional constraints which reduce the space of local solutions of the ``null-curvature'' equations with the Jordan conditions  to the space of local solutions of the generalized Ernst equations are equivalent to the condition of existence for the linear system (\ref{LinSys}) in all cases ($N=2,3,2d$) of the Hermitian matrix integral $\mathbf{K}(w)$ of the form (\ref{KIntegral}) -- (\ref{Omegas}), and for $N=2d$ of additional antisymmetric matrix integral $\mathbf{L}(w)$ of the form (\ref{LIntegral}), which possess, due to the normalization conditions (\ref{NormPsi}), (\ref{Wo}), the expressions (\ref{NormIntegrals})\footnote{We recall here that in the matrix case $N=2d$ we have restricted our consideration by the models with the symmetric $d\times d$ - matrix Ernst potentials. For the models with the Hermitian $d\times d$ - matrix Ernst potentials the integral $\mathbf{K}(w)$ should be symmetric and the integral $\mathbf{L}(w)$ should be Hermitian  \cite{Alekseev:2004}.}.

The holomorphicity of $\mathbf{\Psi}$ and $\mathbf{\Psi}^{-1}$ outside the cut implies that for the conditions  (\ref{NormIntegrals}) to be satisfied, it would be sufficient to satisfy them at the points of the cut. Using in the expressions  (\ref{NormIntegrals}) for these integrals the local representations (\ref{LocStrucA}), (\ref{LocStrucB}) for $\mathbf{\Psi}$ and $\mathbf{\Psi}^{-1}$ on the cuts, one can show that these conditions impose some constraints on the extended monodromy data (\ref{ParametrizationA}). These constraints can be easily solved. In the matrix case ($N=2d$) the condition of existence of the antisymmetric integral $\mathbf{L}(w)$ leads to the condition that the matrices $\mathbf{u}_\pm$ and $\mathbf{p}_\pm$ should be symmetric ones. The condition of existence of the Hermitian integral $\mathbf{K}(w)$ leads to the expression of the components  of  $\mathbf{l}(w)$  in terms of Hermitian conjugated functions $\mathbf{k}^\dagger(w)$:\footnote{It is necessary to take into account here that because of different structures of the cuts in the hyperbolic ($\epsilon=1$) and in the elliptic ($\epsilon=-1$) cases, the action of the defined earlier Hermitian conjugation even on the scalar functions of the spectral parameter is different:
$$\underline{\epsilon=1}:\quad\mathbf{u}^\dagger(w)=
\left\{\begin{array}{ll}
\overline{\mathbf{u}_+^T(\overline{w})},&w\in\Omega_+\\
\overline{\mathbf{u}_-^T(\overline{w})},&w\in\Omega_-
\end{array}\right.\qquad
\underline{\epsilon=-1}:\quad\mathbf{u}^\dagger(w)=
\left\{\begin{array}{ll}
\overline{\mathbf{u}_-^T(\overline{w})},&w\in\Omega_+\\
\overline{\mathbf{u}_+^T(\overline{w})},&w\in\Omega_-
\end{array}\right.
 $$}
\begin{equation}\label{klrelations}
\mathbf{l}(w)=S_0^2(w)\mathbf{W}_0^{-1}(w)\cdot \mathbf{k}{}^\dagger(w)
\end{equation}
where the scalar multiplier $S_0^2(w)=(w-\xi_0)(w-\eta_0)$ is introduced here to cancel the zeros of the determinant of   $\mathbf{W}_0(w)$ and to make the components of $\mathbf{l}_+(w)$ and $\mathbf{l}_-(w)$ holomorphic on the cuts  $L_+$ and $L_-$ respectively.

\paragraph{Monodromy data for solutions of reduced Einstein's field equations}
The constraints (\ref{klrelations}) show that we can chose for a  ``coordinates'' in the space of local solutions of the reduced Einstein's field equations, the affine coordinates of the projective vectors
\begin{equation}\label{ParametrizationB}
\begin{array}{l}
\mathbf{k}^{N=2}(w)=\{1,\mathbf{u}(w)\},\\[1ex]
\mathbf{k}^{N=3}(w)=\{1,\mathbf{u}(w),\mathbf{v}(w)\},
\end{array}\hskip3ex
\left\{\begin{array}{l}
\mathbf{k}^{N=2d}(w)=\{ I_d,\,\,\mathbf{u}(w)\}\\[1ex]
\mathbf{u}^T(w)=\mathbf{u}_(w)
\end{array}
\right.
\end{equation}
only. In these expressions the notations defined in (\ref{Doublefunctions}) and in the footnote to (\ref{klrelations}) are used. Thus, a complete set of independent functional parameters consists of the scalar functions $\mathbf{u}_\pm(w)$ for vacuum gravitational fields ($N=2$), which can be supplied with the functions $\mathbf{v}_\pm(w)$ responsible for the presence of electromagnetic fields ($N=3$) and with the functions $\mu_+(w)$ and $\mu_-(w)$ which determine a massless Weyl spinor field ($N=2,3$), or it consists of $d\times d$ - matrix functions $\mathbf{u}_\pm(w)$ for the string gravity models which we consider here. In the last case $N=2d$, the matrices $\mathbf{u}_\pm(w)$ should be symmetric for the models with symmetric Ernst potentials.\footnote{For the models with Hermitian Ernst potentials, the Hermitian conjugation ${}^\dagger$ in  (\ref{klrelations}) must be changed to the matrix transposition and the matrices $\mathbf{u}_\pm(w)$ in (\ref{ParametrizationB})  must satisfy, instead of the symmetry condition, to the constraint $\mathbf{u}^\dagger(w)=-\mathbf{u}(w)$.}
If we chose instead of generic extended monodromy data  (\ref{ParametrizationA}) the monodromy data of the form  (\ref{klrelations}), (\ref{ParametrizationB}), the space of local solutions of the ``null-curvature'' equations reduces to the space of all local solutions of the corresponding reduced Einstein's field equations.

\paragraph{Calculating solutions of generalized Ernst equations for given monodromy data.} Similarly to the construction of solutions for the ``null-curvature'' equations described above, we have to chose at first some monodromy data of the form (\ref{klrelations}), (\ref{ParametrizationB}) and to find the corresponding solution of one of the basic integral equations (\ref{IntEqs}). Using this solution, we have to calculate then the quadratures (\ref{Rmatrix}) which determine the components of the matrix $\mathbf{R}$ and after that we calculate the corresponding expressions for $\mathbf{U}$ and $\mathbf{V}$ using (\ref{UVR}). The scalar Ernst potentials $\mathcal{E}$ for $N=2$ or $\mathcal{E}$, $\Phi$ for $N=3$ or $d\times d$ - matrix Ernst potentials $\mathcal{E}$ for the case  $N=2d$ can be calculated from the relations $\partial_\xi\mathcal{E}=-\mathbf{U}_{(1)}{}^{(2)}$, $\partial_\eta\mathcal{E}=-\mathbf{V}_{(1)}{}^{(2)}$ and $\partial_\xi\Phi=\mathbf{U}_{(1)}{}^{(3)}$, $\partial_\eta\Phi=\mathbf{V}_{(1)}{}^{(3)}$,
and they are determined by the following expressions:
$$\underline{N=2d}:\hskip1ex\mathcal{E}=\mathcal{E}_0-2 i \mathbf{R}_{(1)}{}^{(2)},\qquad
\underline{N=3}:\hskip1ex\mathcal{E}=\mathcal{E}_0-2 i \mathbf{R}_{(1)}{}^{(2)},\hskip1ex \Phi=2 i \mathbf{R}_{(1)}{}^{(3)},$$
where $\mathcal{E}_0$ is the value of the Ernst potential at the initial point determined by the normalization conditions, and the indices in parenthesis for $N=2,3$ mean the numbers of the matrix components, but for the case $N=2d$ the expression $(\ldots)_{(1)}{}^{(2)}$ means the upper right $d\times d$ - block of a  $2d\times 2d$ - matrix.

\bigskip
Thus, the solution of the direct and inverse problems of the monodromy transform  determines a one-to-one correspondence between the space of local solutions of the ``null-curvature'' equations and the space of arbitrarily chosen extended monodromy data (\ref{ParametrizationA}) as well as the space of solutions of generalized Ernst equations and a free space of the monodromy data (\ref{klrelations}), (\ref{ParametrizationB}). Saying briefly, the monodromy data can play the role of ``coordinates'' in the infinite dimensional space of local solutions of these integrable reductions of Einstein's field equations. The basic system of the linear singular integral equations, which determines this mapping, plays in our approach the role of the dynamical field equations and it can be used effectively for solution of the considered above Einstein's field equations in the presence of space-time symmetries.

\bigskip
The author thanks A.K. Pogrebkov and A.B. Shabat for useful discussions and criticism. This work is supported partly by the Russian Foundation for Basic Research (grants 05-01-00219 and 05-01-00498) and by the programs ``Mathematical Methods of Nonlinear Dynamics'' of Russian Academy of Sciences and ``Leading Scientific Schools'' of Russian Federation (the grant NSh-1697.2003.1).


\begin{thebibliography}{99}\itemsep=0pt

\bibitem{Belinskii-Zakharov:1978} V. A. Belinski and  V. E. Zakharov, Sov. Phys. JETP {\bf 48}, 985, (1978); Sov. Phys. JETP {\bf 50}, 1 (1979).
\bibitem{Harrison:1978} B. K. Harrison, Phys. Rev. Lett.
{\bf 41}, 1197 (1978).
\bibitem{Neugebauer:1979} G. Neugebauer, J. Phys. {\bf A12}, L67
(1979); J. Phys. {\bf A13}, 1737 (1980).
\bibitem{Hauser-Ernst:1979a} I. Hauser and F. J. Ernst,
Phys. Rev. {\bf D20}, 362 (1979).
\bibitem{Kinnersley:1977} W. Kinnersley, J. Math. Phys. {\bf
18}, 1529 (1977).
\bibitem{Kinnersley-Chitre:1977} W. Kinnersley and D. M.
Chitre, J. Math. Phys. {\bf 18}, 1538 (1977).
\bibitem{Kinnersley-Chitre:1978} W. Kinnersley and D. M.
Chitre, J. Math. Phys. {\bf 19}, 1926 (1978).
\bibitem{Hauser-Ernst:1979b} I. Hauser and F. J. Ernst, Phys. Rev. {\bf D20}, 1783 (1979).
\bibitem{Alekseev:1980} G.A. Alekseev,  JETP Lett. {\bf 32} 277 (1980).
\bibitem{Alekseev:1983}  G.A. Alekseev, Sov. Phys. Dokl. (USA) {\bf 28}, 133 (1983).
\bibitem{Belinskii:1979a} V.A. Belinsky, Sov.Phys.JETP {\bf 50}, 623 (1979).
\bibitem{Belinskii:1979b} V.A. Belinsky, JETP Lett. {\bf 30}, 32 (1979).
\bibitem{Bakas:1994} I. Bakas, Nucl. Phys. {\bf B428} (1994) 374;  hep-th/9402016.
\bibitem{Kumar-Ray:1995}  A. Kumar and K. Ray, Phys.Lett. {\bf B358} (1995) 223.
\bibitem{Gal'tsov-et-alii} D.V. Gal'tsov, Phys.Rev.Lett. {\bf 74} (1995) 2863; D. V. Gal'tsov and O. V. Kechkin, Phys. Lett. {\bf B361} 52 (1995); Phys. Rev. {\bf D54} 1656 (1996);  D.V. Gal'tsov and S.A. Sharakin, Phys.Lett. {\bf B399} (1997) 250.
\bibitem{Das-Maharana-Melikyan:2002}  Ashok Das, J. Maharana, A. Melikyan, {\it Monodromy, Duality and Integrability of Two Dimensional String Effective Action}, in proceedings of the Workshop on Integrable Field Theories, Solitons and Duality, Sao Paulo (2002); hep-th/0210012.
\bibitem{Alekseev:2004} G.A.Alekseev, Theor. Math. Phys. {\bf 144}(2) 1065-–1074 (2005); hep-th/0410246.
\bibitem{SKMHH:2003} Stephani, H., Kramer, D., MacCallum, M., Hoenselaers, C. and Herlt, E.,   {\it Exact solutions of Einstein's field equations}, 2nd edition, Cambridge University Press, (2003).
\bibitem{Belinski-Verdaguer:2001}   V. Belinski and E. Verdaguer, {\it Gravitational Solitons}, Cambridge University Press, Cambridge Monographs on Mathematical Physics, (2001).
\bibitem{Alekseev:1987} G. A. Alekseev, Proc. Steklov Inst.
Maths. {\bf 3}, 215 -- 262 (1988).
\bibitem{Sibgatullin:1984} N.R. Sibgatullin, {\it Oscillations and Waves in Strong Gravitational and Electromagnetic Fields}, Nauka, Moscow (1984); English translation: Springer-Verlag, Berlin (1991).
\bibitem{Neugebauer:1981}    G.Neugebauer, Physics Letters {\bf 86A}, no. 2, 91 (1981).
\bibitem{Alekseev:1985} G.A. Alekseev, Sov.Phys.Dokl. {\bf 30}, 565 (1985)
\bibitem{Alekseev:2000} G.A.Alekseev, {\it Monodromy transform approach to solution of some field equations in General Relativity and string heory}, in Proceedings of the workshop "Nonlinearity, Integrability and all that: Twenty years after NEEDS'79",  p. 12 -- 18,  World Scientific, Singapore  (2000); gr-qc/9911045.
\bibitem{Alekseev:1993a} G.A.Alekseev, {\it Explicit form of the extended family of electrovacuum solutions with arbitrary number of parameters}, Abstracts of Contributed Papers, 13th International Conference on General Relativity and gravitation, (Huerta Grande, Cordoba, Argentina),  p. 3 -- 4 (1992).
\bibitem{Alekseev-Garcia:1996} G.A.Alekseev, A.A.Garcia, Phys.Rev. {\bf D53}, 1853 (1996).
\bibitem{Alekseev:1993b} G.A.Alekseev, {\it Integrability of the boundary value problems for the Ernst equations}, Proceedings of
the workshop ``Nonlinear Evolution Equations and Dynamical Systems'', (NEEDS-92, Dubna, 1992), World Scientific, Singapore (1993), p. 5 -- 10.
\bibitem{Ernst:1968} F.J.Ernst, Phys.Rev. {\bf 167} (2), 1175; {\bf 168} (2), 1415 (1968).
\bibitem{Hauser-Ernst:1980} I. Hauser and F. J. Ernst, J. Math. Phys. 21, 1126 (1980).
\bibitem{Hauser-Ernst:2001} I. Hauser and F. J. Ernst,
Gen. Rel. Grav., {\bf 33}, 195 (2001).
\bibitem{Muskhelishvili:1968} N.I.Muskhelishvili, {\it Singular Integral Equations}, 3rd ed., Nauka, Moscow (1968); English translation of 1st ed.: Noordhoff, 1953; reprinted (1972).
\bibitem{Gakhov:1977} F.D. Gakhov, {\it Boundary Value Problems}, 3rd ed., Nauka, Moscow (1977); English translation of 2nd ed.: Pergamon, Oxford, and Addison-Wesley, Reading, MA (1966).

\end{thebibliography}
\end{document}